\documentclass[12pt,preprint]{aastex}

\def \eg           {{e.g.}}
\def \etal         {{et~al. }}
\def \ie           {{i.e.}}

\def \ha           {\hbox{H$\alpha$}}
\def \kms          {\hbox{km$\,$s$^{-1}$}}

\def\approxlt{\lower.2em\hbox{$\buildrel < \over \sim$}}
\def\approxgt{\lower.2em\hbox{$\buildrel > \over \sim$}}

\def \lir          {\hbox{$L_{\mbox{\tiny IR}}$}}
\def \ls           {\hbox{L$_{\odot}$}}
\def \ms           {\hbox{M$_{\odot}$}}           

\def\kms{\ifmmode{\,\hbox{km}\,s^{-1}}\else {\rm\,km\,s$^{-1}$}\fi}

\def\ms{{\rm\,M_\odot}}
\def\lsun{{\rm\,L_\odot}}

\def\cm{{\rm\,cm}}

\def\kmps{{\rm\,km\,s$^{-1}$}~}
\def\kmpsnospace{{\rm\,km\,s$^{-1}$}}
\def\hmpc{\ifmmode{h^{-1}\,\hbox{Mpc}}\else{$h^{-1}$\thinspace Mpc}\fi}

\def\eg{{\it e.g.}~}
\def\etal{{\it et~al.}~}

\def\ie{{\it i.e.}~}

\def\spose#1{\hbox to 0pt{#1\hss}}
\def\lta{\mathrel{\spose{\lower 3pt\hbox{$\mathchar"218$}}
\raise 2.0pt\hbox{$\mathchar"13C$}}}
\def\gta{\mathrel{\spose{\lower 3pt\hbox{$\mathchar"218$}}
\raise 2.0pt\hbox{$\mathchar"13E$}}}

\def\H2{$H_2$~}
\def\H2S{$H_2^*$~}
\def\Mco{{$M_{H_2}$~}}

\def\D25{{$D_{25}$~}}

\def\McoHI{{$M_{H_2}/M_{HI}$~}}

\def\HIH2LB{{$M_{GAS^*}/L_B$~}}

\def\SFE{{$L_{IR}/M_{H_2}$~}}

\def\MH2D2{{$M_{H_2}/D_{25}^2$~}}
\def\MH2D2S{{$M_{H_2^*}/D_{25}^2$~}}

\def\thCO{{$^{13}$CO~}}
\def\TA{{$T_A^*$}}
\def\Tk{{$T_k$~}}
\def\XdVunit{{\rm\,pc\,{[km\,$s^{-1}$]}$^{-1}$}~}
\def\Xunit{{\rm\,cm$^{-2}$\,{[K km\,$s^{-1}$]}$^{-1}$}~}

\def\1213CO{{[$^{12}$CO}/{$^{13}$CO]~}}
\def\nh2{{$n(H_2)$}}
\def\cm3{{$cm^{-3}$}}
\def\H2{{$H_2$}}

\begin{document}
\title{GAS AND DUST IN THE TAFFY GALAXIES: UGC~12914/15}

\author{Ming Zhu}
\affil{Joint Astronomy Centre, National Research Council Canada, 
660 N. A'ohoku Place, Hilo, Hawaii 96720, USA}
\author{Yu Gao}
\affil{Purple Mountain Observatory, Chinese Academy of Sciences, 2 West
Beijing Road, Nanjing 210008, China}
\author{E. R. Seaquist}
\affil{University of Toronto, Department of Astronomy and
Astrophysics,  60 St. George Street, Toronto, ON, M5S 3H8,
Canada}
\author{Loretta Dunne}
\affil{Institute for Astronomy, Royal Observatory Edinburgh, 
Blackford Hill, Edinburgh, EH9 3HJ, UK}

\begin{abstract}

We present a comprehensive study of the dust and gas properties in
the after-head-on-collision UGC~12914/15 galaxy system using
multi-transition CO data and SCUBA sub-mm continuum images at
both 450 and 850$\mu$m. CO(3-2) line emission was detected in the disks
of UGC 12914 and UGC 12915 as well as in a bridge connecting the
two galaxies. Dust emission at 450$\mu$m was detected for the first time
in the two galactic disks and in the connecting bridge. Using an LVG
excitation analysis model we have obtained good estimates of the
physical parameters in different regions of this system and the
amount of molecular gas was found to be 3--4 times lower than that
estimated by other investigators using the standard Galactic
CO-to-H$_2$ conversion factor. Comparing with the dust mass derived
from the SCUBA data, we found that the gas-to-dust ratio was
comparable to the Galactic value in the two galaxy disks but a
factor of $\sim 3$ higher in the bridge. The physical condition of the
molecular gas in the bridge is comparable to that in the diffuse
clouds in our Galaxy. Our result is consistent with the scenario
that the bridge molecular gas originated from the disk molecular
clouds and has been
drawn out of the galactic disks due to direct cloud-cloud collision.

Our data indicate that the global star formation efficiency (SFE, \SFE) in
UGC 12915 is comparable to that of normal spiral galaxies, and the
SFE is 40\% lower in UGC 12914 than in UGC 12915. Little star
formation activity was found in the bridge except in an \ion{H}{2} region
adjacent to the disk of UGC 12915.

\end{abstract}


\section{INTRODUCTION}

Galaxy interactions play a crucial role in galaxy evolution,
especially in the early universe where collisions and interactions are
expected to be more frequent.  It has been established that slow speed
close collisions which lead to the merging of two galaxies could
trigger intense starbursts. Gas-rich mergers could reach the phase of
an ultraluminous infrared galaxy, and
eventually lead to the formation of an elliptical galaxy (Sanders \&
Mirabel 1996). However, not all galaxy interactions will result in a
merger and it is equally important to study different types of galaxy
interaction to understand their role in galaxy evolution.

The Taffy galaxy pair (UGC 12914/15) represents a post collision
system consisting of two counter-rotating disks with a remarkable
bridge of gas and dust between them (see the background image in Fig
2). The southern galaxy, UGC 12914, contains a stellar ring which
suggests that it has suffered a direct hit by the intruder, UGC 12915,
at a speed much higher than the galactic rotation velocities (Struck
1999). During the ``splash'', collisions between stars are unlikely,
but collision between the interstellar medium (ISM) of the two disks
is inevitable.  Previous CO and \ion{H}{1} observations have revealed
huge amounts of atomic and molecular gas between the two galactic
disks (Condon et al. 1993; Gao, Zhu \& Seaquist 2003, hereafter
GZS03). This gas is thought to be either dragged out from the disks, or the
result of direct cancellation of the motion of the giant molecular
clouds (GMCs) in the disks after the counter-rotating,
interpenetrating and head-on collision.  Condon et al. (1993)
dubbed this system the ``Taffy'' galaxies when they discovered a very
bright radio synchrotron bridge connecting the two spiral galaxies in
the VLA 20 cm continuum map. The \ion{H}{1} flux density peaks at the
center of the bridge, and both single dish (Braine et al. 2003) and
interferometer measurements (GZS03) of the CO(1--0) show that there
are $4-14 \times 10^9 \ms$ of molecular gas in the bridge region
(assuming a conventional $X$ factor), more than the total amount of
that in our Galaxy. The origin of this huge amount of gas and its
fate is crucial to understanding the ISM evolution and star formation
in this interacting pair.

The Taffy galaxy pair is very luminous in far-infrared (FIR), with
a total \Mco $= 4.5 \times 10^{10} \ms $ and $\lir= 8.1 \times 10^{10} $ \ls
(GZS03), essentially qualifying as a luminous IRAS galaxy.  The mid-IR
(MIR) image (Jarrett et al. 1999) from the Infrared Space Observatory
(ISO) shows that UGC 12915 is very bright in the MIR, suggesting
vigorous star formation activity in this system. Outside the disk of
UGC 12915, an \ion{H}{2} region is clearly seen in the ISO image as
well as in the $H_\alpha$ image (Bushouse \& Werner 1990), suggesting
that stars are forming in the bridge.  However, the global star
formation efficiency (SFE), \lir/\Mco is $\sim 2 $ $\ls/\ms$ is
relatively low compared to many normal spiral galaxies, and 10 times
lower than that of most luminous IRAS galaxies (Gao \& Solomon 1999).
Why is the global SFE so low for such a strongly disturbed gas rich
system? Has the mass of molecular gas been
overestimated by using a standard conventional CO-to-H$_2$ conversion
factor $X$?

In this paper, we present a comprehensive study of the dust and
molecular gas in the Taffy galaxies. We have mapped the cold dust
emission at both 450 and 850$\mu$m using SCUBA on the James Clerk
Maxwell Telescope (JCMT), which allows us to make an independent
estimate of the gas mass.  We have also used the JCMT to observe the
CO(3--2) and CO(2--1) line emission in various regions. Combining
these data with the CO(1--0) data from the BIMA interferometer (GZS03)
and the CO(2--1) data from the IRAM 30m telescope (Braine et
al. 2003), we have used an LVG model to constrain the physical
parameters in the molecular clouds. This allows us to make a more
accurate estimate of the $X$ factor and hence a more reliable
measurement of the molecular gas mass in the bridge.

We assume a Hubble constant $H_0=75 $ km s$^{-1}$ Mpc$^{-1}$
throughout. The distance is assumed to be 59.3 Mpc in our
calculations.

\section{Observations and data reduction}

\subsection{SCUBA Observations}

Simultaneous observations at 450 and 850$\mu$m using the Submillimetre
Common-User Bolometer Array (SCUBA) were carried out at the JCMT on 23
Dec 2002 and 16 Sep 2004 in exceptionally good weather with CSO
Tau=0.03 at 225 GHz. Two pointings were made centered on the two
galaxy nuclei in order to make a mosaic image of the entire system.
These two pointings provide sufficient coverage of the taffy feature
between the two galaxies and the resulting best rms is in the overlap field
of the mosaic. There is full coverage of the whole system including the
tidal features/tails which are prominent in the
\ion{H}{1} maps and deeper optical images. The Taffy galaxies were
also observed in 1998 as part of the SLUGS survey (Dunne et
al. 2000). We have retrieved these data from the JCMT archive and
combined all of them to make an image at 850$\mu$m.  The 450$\mu$m map
contains only data from 2002 and 2004 because the 1998 data were taken
in poorer atmospheric conditions and were therefore not useful at this
wavelength. The 450$\mu$m data are of exceptional quality and show the
value of working at this wavelength when conditions allow. Uranus and
CRL618 were used for flux calibration.  Different flux conversion
factors (FCFs) have been applied to the raw data before
co-adding. Table 1 summarizes the observational details.

Pointing was checked regularly by observing the nearby blazar 2251+158
and was always better than $3''$, much smaller than the primary beam
of the array of $8''$ at 450 $\mu$m and $15''$ at 850 $\mu$m.  The
average sky opacity in terms of CSO $\tau$ at 225 $\mu$m during the
observations is given in column 6 of Table 1.

All of the maps were made in jiggle-mode where the chop throw and
direction were chosen to avoid chopping onto emission from the
neighbouring galaxy. The data were reduced using the standard SURF
reduction routines (Jenness \& Lightfoot 1998), following the steps:
flat-fielding, extinction correction using the polynomial fits to the
CSO tau data, spike removal and noisy bolometer removal prior to sky
subtraction. To remove the sky the task 'remsky' was used and the
bolometers to estimate the sky level were chosen to avoid the region
containing the galaxies and bridge (where source emission was likely
to be present). The median method was chosen to remove the sky and the
subtracted value not added back on. The individual sub-maps were
corrected for their non-zero baselines before co-adding and the noise
on each map used to determine its weight in the final co-add. The data
sets were co-added and regridded using 'rebin'.

A noise map was also made using a Monte-Carlo method (described in
detail in Eales et al. 2000). This allows us to account for the
varying noise levels across the image, which are rather non-uniform
due to data being taken on different nights and also because of noisy
pixels. The sub-mm images shown in Fig. 1 are signal-noise images
made from the ratio of the flux map with the noise map.

We estimate a conservative uncertainty of 15\% and 30\% in the derived
fluxes at 850$\mu$m and 450$\mu$m respectively. The statistical
uncertainty on the fluxes was estimated using the formula given by
Dunne et al. (2000) and found to be of order 5-9\% at 450 and
850$\mu$m. The main source of uncertainty in the final fluxes is
therefore the absolute calibration.  The rms is 11.8 mJy/beam at
450$\mu$m and 3.3 mJy/beam at 850$\mu$m in the overlap region of the
mosaic image and slightly higher in the outer regions.

\subsection{ The CO(3--2) and (2--1) Observations}

Half of the $ ^{12}$CO(3--2) data were taken in one observing run
during Aug 1999. The other $ ^{12}$CO(3--2), (2--1) and $
^{13}$CO(2--1) data were taken in flexible scheduling mode at the JCMT
during Aug and Oct 2003.  Receiver B3 was used to
observe the $ ^{12}$CO and \thCO J=3--2 transitions at 345.796 GHz,
while receiver A2 was used to observe the $^{12}$CO and \thCO(2--1)
transitions at 230.538 GHz and 220.399 GHz respectively.  The B3
receiver has dual polarized mixers, permitting an increase in
sensitivity by a factor of $\sqrt{2}$ when observing in the
dual-channel mode. The system temperatures measured were $\rm T_{\rm
sys}\sim 300-500$ K for receiver B3 and $\rm T_{\rm sys}\sim 300$ K
for receiver A2.

The Digital Autocorrelation Spectrometer (DAS) was employed which
provides a total usable bandwidth of 920 MHz (802\kmpsnospace) for B3
and 760 MHz (980 \kmpsnospace) for A2.  The spectrometer band was
centered at $\rm V_{\rm LSR} = 4400$ {\rm\,km\,s$^{-1}$}.  The
frequency resolution was 0.625~MHz, which corresponds to a velocity
resolution of 0.545 \kmps at 345 GHz and 0.84 \kmps at 230 GHz.  The
beam shape of the telescope is closely approximated by a Gaussian with
HPBW=14$''$ at 345 GHz and 20$''$ at 230GHz.

All the data were taken by beam-switching in azimuth using a beam
throw of 150$''$ at PA=90.  Pointing was monitored frequently by
observing the blazar 2251+158 and the variable star RAnd
(spectral-line pointing).  The rms pointing error was found to be
$\sim 2.5''$ for CO(3--2) and $\sim 3''$ for CO(2--1).  The spectral
line intensity is calibrated in terms of \TA, \ie, corrected for
atmospheric transmission and losses associated with telescope
efficiencies and rearward scatter. Conversion to $T_{\rm mb}$ was done
by using the formula $T_{\rm mb} = T_A^*/\eta_{mb} $, where the main
beam efficiency $\eta_{mb}$ is $0.62 \pm 0.05 $ for B3 and $0.69 \pm
0.08 $ for A2 according to the measurements of Uranus and Mars.

Measurements of the CO(3--2) spectral line of CRL618, OH231.8,
NGC6334I at $\rm V_{\rm LSR}=0$ km s$^{-1}$ were used to estimate a
calibration uncertainty of $\sim 10$\% by comparing with standard
spectra in the JCMT archives. Finally, we repeatedly monitored the
spectral line emission of the map center throughout our runs and found
excellent agreement among all the line profile measurements.

With limited telescope time, we mapped the Taffy galaxies at selected
positions where the CO emission is strong enough for a quantitative
study.  The central region of UGC 12915 was fully sampled with $7''$
spacing.  Other points were chosen to match the IRAM data at
the CO(2--1) transition. The integration time per point is typically
10--30 minutes depending on the strength of the signal.

Most of the CO(2--1) data were observed using the IRAM 30-m
telescope. The observational details were described in Zhu et
al. (1999).  These data cover the entire system, but with $11''$
spacing.  We use the JCMT to re-observe the CO(2--1) line at selected
positions in order to calibrate the fluxes measured from the different
telescopes with different resolutions.

The JCMT software SPECX was used to reduce the spectral data and a
linear baseline was removed from each spectrum. Table 2 lists the
integrated intensity of CO(3--2) and the results are discussed in \S
3.3.

\section{Result and Analysis}

\subsection{ Dust Emission}

Figure 1 shows the 850$\mu$m contours overlaid on an 450$\mu$m color
map.  The sub-mm emission is mainly in the two galactic disks, but a dust
bridge is clearly detected at both 450 and 850$\mu$m. The S/N is
actually higher on the 450$\mu$m map, and this is due to a combination
of the superb weather and the rather poor behaviour of the 850$\mu$m
array, which contained many noisy bolometers. The lack of redundancy at
850$\mu$m (having on 37 pixels rather than 91 as at 450$\mu$m) means
that some areas of the image have less data than others and these 'bad'
regions often fell on the galaxy disks or bridge region. For this
reason, the detailed morphology of the emission at 850$\mu$m is less
reliable than at 450$\mu$m. This should be borne in mind by the reader
when there appear to be differences in the positions of the peaks between
the 850$\mu$m, 450$\mu$m and CO maps. In general, however, the maps from the
two SCUBA wavelengths are consistent with each other.

Figure 2 presents the 450$\mu$m contours overlaid on a MIR 8$\mu$m
image from the Spitzer Infrared Array Camera (IRAC).  The dust
emission from SCUBA is quite compact in the inner disk of UGC 12915,
but rather evenly distributed between the two spiral arms and the
nuclear region in UGC 12914. The cold dust seems to extend out along
the both spiral arms into the ring in UGC 12914, suggesting that tidal
forces have also affected the distribution of the cold dust. In the
bridge, a significant fraction of the 450 and 850$\mu$m fluxes are associated
with the \ion{H}{2} region outside the disk of UGC 12915.  Further
away from the \ion{H}{2} region, the emission seems to be associated
with the ring of UGC 12914, but it is difficult to distinguish the
bridge from the ring due to the low resolution of SCUBA. The
\ion{H}{2} region appears to be at the edge of the sub-mm peak at the 
bridge, suggesting that it might be just emerging from the dusty clouds and
starting to clear away the nearby cold dust.

\subsubsection{Comparison with hot dust}

To compare with the hot dust distribution, we overlay the 450$\mu$m
  contours on a mid-IR image at 15$\mu$m from ISO (Jarrett et
  al. 1999, Fig. 3). The ISO image is in logarithmic scale and has
  been contrasted to show the low surface brightness emission. Both
  the Spitzer 8$\mu$m image and the ISO 15$\mu$m image trace the hot
  dust. The source of the MIR emission is thought to be polycyclic
  aromatic hydrocarbon (PAH) molecules or tiny carbonaceous grains
  heated by UV photons. The bridge is more apparent in the ISO
  15$\mu$m image, while the Spitzer image has higher resolution and
  better shows the features seen in optical wavelengths, i.e., the two
  galactic disks, the ring of UGC 12914 and the \ion{H}{2} region
  outside the disk of UGC 12915.

The morphology of the cold dust distribution traced by the 450$\mu$m
and 850$\mu$m emission is basically consistent with that of the hot
dust.  Table 3 lists the measured sub-mm flux densities in different
regions, along with the ISO 15$\mu$m, CO and
\ion{H}{1} fluxes (the comparison with CO and \ion{H}{1} fluxes will
be discussed in next section).  The bridge contains approximately 18\%
of the total 850$\mu$m flux, yet less than 2\% of the 15 $\mu$m
flux. The northern galaxy (UGC 12915) contains 41\% and 44\% of the
850 and 450$\mu$m fluxes, while the UGC 12914 disk contains 41\% and
36\% at these two wavelengths. The flux ratio $S_{450}/S_{850}$ is
rather constant across the entire system (at least on a global scale),
suggesting that the properties of the cool dust component are similar
in the bridge and the two galaxies. Breaking down further to
individual regions, we have measured the fluxes in a $25''$ aperture
centering at the nuclei of the two galaxies and the southern and
northern MIR knots of UGC 12914 (Table 3). We can see that the flux
ratio $S_{450}/S_{850}$ in the disk of UGC 12914 is slightly lower in the
northwest than the southeast, suggesting that the dust is colder in
the northern disk of UGC 12914. 
This is consistent with the finding
of Jarrett et al. (1999) based on the ISO MIR fluxes, in which they
found that the MIR emission in northern UGC 12914 is weaker and redder
in terms of the MIR color [11.4 $\mu$m]/[15 $\mu$m].  Since the MIR
emission is produced by hot dust heated by young stars, it is a good
tracer of star formation sites. The disk of UGC 12915 contributes 61\%
of the MIR flux, presumably generated by vigorous starbursts, hence
the gas and dust is expected to be hotter in UGC 12915.

\subsubsection{ Comparison with Molecular and Atomic Gas Distribution}

Figs 4a and 4b present the 450 and 850$\mu$m contours overlaid on a
CO(1--0) image from the BIMA interferometer, while Fig 5 shows the
850$\mu$m contours overlaid on an \ion{H}{1} image from the VLA.  Gao
et al. (GZS03) have made a thorough comparison between the CO and the
\ion{H}{1}, MIR and 20 cm radio continuum data. The bridge is very
prominent in the \ion{H}{1}, CO and SCUBA 450 $\mu$m map, but is less
conspicuous in the 850 $\mu$m map (less than 3$\sigma$ in most bridge
region).  At 850$\mu$m, the 4$\sigma$ contours looks more like a tail
extended out from the \ion{H}{2} region, rather than a bridge
connecting the two disks. The width of this extension is $\sim 16''$,
barely resolved by the $15''$ beam.  However, there appear to be some
lower surface brightness structures between the two galactic disks at
2-3$\sigma$ level. We can estimate whether this is consistent
with the CO(1--0) data. At the southwest end of the bridge, for
example, at offset ($-22''$, $-22''$) from the nucleus of UGC 12915,
the CO(1--0) integrated intensity $I_{CO}$ is 21 Jy K km s$^{-1}$ in
the BIMA map (GZS03), which is 1/8 of the peak flux in UGC 12915. If
the $S_{850}/I_{CO}$ flux ratio is constant, the dust emission should
be 1/8 of the peak 850 $\mu$m fluxes, which would be 7 mJy and
therefore would yield only a 2$\sigma$ detection.  These weak
structures are seen at both 450 and 850 $\mu$m, suggesting we are
seeing a faint dusty bridge similar to that of the CO.

However, the peak sub-mm emission in the bridge is slightly offset
from that of the CO and appears to be located between the CO and
\ion{H}{1} peaks (Fig. 4 and 5).  This suggests that part of the dust
may be associated with \ion{H}{1}.  A similar trend is also seen in
the northern spiral arm of UGC 12914, in which some of the 850 $\mu$m
continuum appears to be associated with the strong \ion{H}{1} emission
in the bridge. The highest \ion{H}{1} concentrations of the entire
system are in the bridge (Fig. 3 in Gao et al. 2003), connecting the
northwestern part of the UGC 12914 disk with the \ion{H}{1} peak, the
\ion{H}{2} region, and the middle of the UGC 12195 disk. CO emission
is essentially the lowest at the \ion{H}{1} peak, but diffuse CO
emission is abundant over this area. The molecular bridge is mostly
related to UGC 12915, connecting the nuclear regions of both disks,
with huge CO concentrations at the \ion{H}{2} region. It is
interesting to see that the cold dust bridge appears to be situated
right between the \ion{H}{2} regions and the \ion{H}{1} peaks, forming
a contiguous mixed bridge connecting the CO, dust, and \ion{H}{1}
peaks between the two galaxies.  The shift between the dust and CO
peaks is only $7''- 8''$ and for comparison the beam size at 850
$\mu$m is 15$''$. The bridge peak is not well defined, with a S/N of
only 5--6. High resolution and high S/N data are needed to better
define the dust bridge and confirm the shift from the CO bridge.

In general though, the dust emission at 450 and 850$\mu$m
generally follows that of the CO, indicating that they are tracing the
same component of the ISM.  The spatial correlation between the dust
and \ion{H}{1} emission is weaker, suggesting that the majority of the
cold dust is associated with dense molecular clouds.

Comparing the CO and sub-mm fluxes in Table 3, we can see that the
flux ratio of $S_{CO}/S_{850}$ is almost a constant across the
system. There is not much difference between the two galaxy disks,
and not much variation within the disk of UGC 12914. Even the bridge
has a similar $S_{CO}/S_{850}$ ratio to the disks. Given that a large
fraction of the 850$\mu$m flux is contamination by the CO(3-2) line
(see section~\ref{dustmass}), this constant ratio may not seem
surprising. However, we also compare the CO flux with that at
450$\mu$m ($S_{CO}/S_{450}$ in Table 3), and here too the ratio is
remarkably similar across the system suggesting that the dust and
molecular gas are well mixed. 

Braine et al. (2003) suggested that dust destruction by shocks could
result in a higher CO/dust mass ratio in the bridge. Our direct flux
measurements do not support this speculation. However we can estimate
the CO-to-H$_2$ conversion factor $X$ for the different regions,
and thus we can check whether this lack of variation in the flux ratios extends
to the derived masses of molecular gas and dust.

\subsection{\label{dustmass}Dust Temperature and Mass}

To estimate the dust temperature and mass, we need to remove the
CO(3--2) contamination to the 850 $\mu$m fluxes in the SCUBA band.
Using the formula in Seaquist et al. (2004), but with the wide band
filter profile from the JCMT website
\footnote{http://www.jach.hawaii.edu/JCMT/continuum/background/scuba\_850filter.txt},
we derived that the SCUBA equivalent flux for the CO(3--2) line is
$S_{CO} = 0.53 \,I_{CO}$ mJy beam$^{-1}$ (K km/s)$^{-1}$, where
$I_{CO}$ is the CO(3--2) integrated intensity in a 15$''$ beam on the
main-beam temperature scale.

Table 2 lists $I_{CO}$ at $14''$ resolution and so we have converted
them to $15''$ to compare with the SCUBA fluxes at each position. We
estimate that about 30\%--40\% of the 850$\mu$m flux is from the
CO(3--2) line and this factor does not vary significantly across the system
(within 20\% which is less than the absolute calibration
uncertainty). Hence we can correct for the CO(3--2) contamination by
applying a factor of 65\% to the measured 850 $\mu$m fluxes and derive
the true 850 $\mu$m fluxes free from the CO(3--2) line emission. The
contribution of CO line emission to the 450$\mu$m fluxes has been
shown to be negligible by Seaquist et al. (2004) and so no correction
is made to the 450$\mu$m flux.

Combining the IRAS fluxes at 60 and 100$\mu$m from the revised IRAS
BGS (Sanders et al. 2003) and the SCUBA fluxes at 450 and 850$\mu$m,
we have derived the total dust mass by modeling the spectral energy
distribution (SED).  We found that the best fit for a single component
model was with $\beta=1.4$, $T_{\rm d} \sim 30 $ K, $M_{\rm dust} \sim
4.5 \times 10^7\ms$, however the fit is poor with a $\chi^2$ of 2.2
(Fig. 6a). Dunne \& Eales (2001) found that galaxies with better
sampled SEDs (including more data points from ISO and/or other
measurements) require a second colder dust component at around 20K. In
addition, a fixed value of $\beta=2$ was strongly suggested by the
uniformity of the $S_{450}/S_{850}$ flux ratio. The need for two
components is also physically sound, as galaxies always contain both
hot dust in star forming regions and cold dust in quiescent
regions. In reality, there will be dust at a range of temperatures
depending on its environment, but two components is the simplest
physically motivated model to fit to the data. Therefore, we also fit
the fluxes with a two component SED using $\beta$ = 2 (Fig. 6b). The
fit is improved with a $\chi^2$ of 0.82. The formula we used is
similar to that of Dunne \& Eales (2001):
\begin{equation}
S(\lambda) = M_{\rm w} B(\lambda ,T_{\rm w}) Q_{\rm em}(\lambda) + M_{\rm c}
B(\lambda ,T_{\rm c}) Q_{\rm em}(\lambda)
\end{equation}
where $S$ is the flux density at wavelength $\lambda$, $M_{\rm w}$ and
$M_{\rm c }$ are the masses of the warm and cold dust component,
$B(\lambda,T)$ is the Planck function, $T_{\rm w}$ and $T_{\rm c}$ are
the warm and cold dust temperatures and $Q_{\rm em}$ is the
wavelength-dependent emissivity of the grains which varies as
$\lambda^{- \beta}$ (here $\beta$ = 2) over the wavelength range
considered. We have assumed a value for $\kappa$ (the dust mass
opacity coefficient (which is related to $Q_{em}$ as $\kappa = 4a\rho /
3Q_{em}$) of 0.077 m$^2$ kg$^{-1}$ at 850$\mu$m, and 0.275 m$^2$
kg$^{-1}$ at 450$\mu$m.

Table 4 lists the results from the two-component SED fitting.  The
total dust mass is 45\% higher than that derived from a one component
model. This difference is due to the colder mean mass-weighted dust
temperature which results from the two component fit and is consistent
with the findings of Dunne \& Eales (2001) for a larger sample of IRAS
galaxies. Given that the two-component SED is a better fit to the data
our discussion will use the dust mass derived from the two component
model.

To derive the dust masses in the individual galaxies, we need to
estimate their individual FIR fluxes. The Taffy galaxies were just
resolved by IRAS at 100$\mu$m and 60$\mu$m in the IRAS High Resolution
Image Restoration (HIRES) Atlas (Surace et al. 2004). Additionally,
Zink et al. (2000) observed the northern galaxy, UGC 12915, twice at
100$\mu$m using the Kuiper Airborn Observatory (KAO) with a resolution
of $36''$, yielding a flux density of 9.9 and 11.6 Jy in two
measurements. We adopt the average value of 10.6 Jy for UGC 12915,
which is consistent with the HIRES flux when scaled to the total flux
from the revised IRAS BGS (Sanders et al. 2003). The KAO observations
did not detect obvious 100$\mu$m emission at the bridge, and did not
cover UGC 12914. The total 100$\mu$m flux for the whole system is 14.1
Jy from IRAS, thus the flux from UGC 12914 should be 3.5 Jy. At
60$\mu$m, we use the ratio given by Surace et al. (2004) for the two
galaxies but scale the absolute fluxes to the total given by Sanders
et al. (2003) as this is considered to be the most reliable. Table 4
lists the IRAS fluxes used. Using these data, we can fit the SED of
UGC 12915 and UGC 12914 separately with a two-component model. The
results are summarized in Table 4 and shown in Figs 6c \& d. For
comparison, we also list the mass of the molecular and atomic gas
components in each region in Table 7. 

The SEDs of both galaxies are dominated by the colder dust component,
but the cold dust is at a markedly higher temperature in UGC 12915 (24
K compared to 18 K in UGC 12914). This probably reflects the more
intense interstellar radiation field in UGC 12915 as a result of its
more vigorous star formation activity. Note, however, that despite UGC
12915 having a factor $>2$ greater $L_{IR}$ and IRAS fluxes and a
comparable 850$\mu$m flux, it has a lower dust mass than UGC
12914. The reason for the difference in mass is that the dust is
colder in UGC 12194, thus more dust is required to produce the same
flux at 850$\mu$m. This relative distribution of dust between the two
galaxies is different from the molecular gas traced by CO (which is
roughly equal in both), but consistent with the distribution of
\ion{H}{1}. UGC 12914 has 20\% more
\ion{H}{1} than UGC 12915, but the total gas-to-dust ratio = ($M_{HI} +
M_{H_2})/M_{\rm dust}$ is similar between UGC 12915 and UGC 12914 if
the $X$ factor is the same (see
\S 3.6).

The total dust mass derived from fitting two galaxies separately is $
7.2 \times 10^7 \ms$, which is consistent with the value $6.9 \times
10^7 \ms$ from fitting the whole system. Given the additional
uncertainties in breaking down the FIR fluxes on an individual basis
(mostly due to the bridge), we will use the value from fitting
the system as a whole, $6.9 \times 10^7 \ms$, as the total dust mass.


\subsection{The CO(3--2) Data and Ratios of the Integrated Line Intensities}

CO(3--2) data have been obtained at selected regions. Figure 7
presents the CO(3--2) profiles, which were overlaid on the BIMA
CO(1--0) zero-momentum color map. Figure 8 shows the CO(3--2) profiles
overlaid on the BIMA CO(1--0) spectra at the three positions listed in
Table 5 for excitation analysis.

The CO(3--2) emission is strong in the central region of UGC 12915 and we
have fully sampled it. These profiles are consistent with that of the
CO(1--0) (see GZS03) and the CO(2--1) (Braine et
al. 2003). The peak velocity is blue shifted at the southeast end of
the UGC 12915 disk and red-shifted at the northwest
end. This is the typical kinematic feature for a rotational disk and
has been discussed in details in Gao et al. (GZS03).

For UGC 12914, the quality of the IRAM CO(2--1) line data was poor at
the northern disk due to unstable weather conditions, so we took only
two CO(3--2) spectra using the JCMT in the southern disk where the CO
emission was found to be strong enough for a quantitative study.

CO(3--2) is also clearly detected in the bridge. With a resolution of
$14''$, the single dish CO(3--2) data reveal large amounts of excited
molecular gas located outside the two galactic disks and some of them
are unarguably separated from the \ion{H}{2} region.  Near the
\ion{H}{2} region, the CO(3--2) emission is even stronger than that on
the disk of UGC 12914.  The CO(3--2) integrated intensity is listed in
Table 2, together with the CO(3--2)/CO(1--0) line intensity ratios.


We define the ratios of integrated intensity of
main-beam temperature as follows:

$$r_{21} = I(^{12}CO(2-1)) / I(^{12}CO(1-0)) $$
$$r_{31} = I(^{12}CO(3-2)) / I(^{12}CO(1-0))$$
$$R_{10} = I(^{12}CO(1-0)) / I(^{13}CO(1-0))$$
$$R_{21} = I(^{12}CO(2-1)) / I(^{13}CO(2-1))$$

The BIMA CO(1--0) map (GZS03) was made using the B,C and D configuration of the
array. The shortest baseline was 6 m which is smaller than the
diameter of an individual dish, thus in theory no short spacing should be
missing and all the CO(1--0) flux should be recovered. In fact, the
CO flux reported by Braine et al. (2003) using the IRAM 30m single
dish telescope is consistent with that measured independently by 
GZS03 from the BIMA map. We have also double checked this by
convolving the BIMA data to $22''$ resolution to compare with the
IRAM CO(1--0) data at a variety of  regions. No missing flux was
found on the disks of UGC 12915 and UGC 12914, and a maximum of 30\% flux
is missing in the bridge region. Thus we can convolve the BIMA data
to a resolution of $14''$ to derive the CO(3--2)/(1--0) ratio $r_{31}$.

Table 2 lists the derived $r_{31}$ values.  The uncertainty is mainly
from CO(3--2), for which the baseline is not easy to determine because
the profiles are rather broad and in some regions occupy almost the
whole bandpass of the spectrometer. The total uncertainty is
estimated to be within 20\%. Pointing errors from both telesopes are much
less than half the beam size, thus are not significant.  The
uncertainty is high at the offset ($7''$,$-7''$) and ($-7''$,$-7''$)
due to poor baseline in the CO(3--2) data.

From Table 2,  we can see that $r_{31}$ 
does not vary significantly across the system.
It is almost a constant of $0.3 - 0.4$, except in
 the bridge region where the ratio is as high as $0.5 - 0.6$
which could be due to missing flux in the CO(1--0) data.

Such  $r_{31}$ values are low compared to the average $\langle r_{31} \rangle = 0.64$ in
the nuclei of nearby spiral galaxies (Devereux et al. 1994) and
IR-luminous galaxies  (Yao et al.  2003). Radiative transfer
analysis shows that $r_{31}< 0.4$ indicates that
the gas is either very cold or of low density  $n < 10^4 $ \cm3. Yao et
al. (2003) show that $r_{31}$ is correlated with the SFE. The low $r_{31}$ in
UGC 12914/15 is consistent with the low SFE in this system.

We have measured $r_{21}$ with $11''$ resolution using the CO(1--0)
data from BIMA and the CO(2--1) data from the IRAM 30-m. The CO(2--1)
data (Braine et al. 2003) have a higher uncertainty than the CO(3--2)
data due to poor weather conditions. Also the bandwidth of the IRAM
receiver at 230GHz was not wide enough to cover the broad line in some
regions, resulting in poorly determined baselines.  Nevertheless, the
available data suggest that $r_{21}$ also shows little fluctuation
across the system.  To evaluate the variation of $r_{21}$ and
$R_{21}$ with different resolutions, we have re-observed the
$^{13}$CO(2--1) and $^{12}$CO (2--1) lines with the JCMT at the three
regions listed in Table 5, with a resolution of $20''$. We found that
$r_{21}$ and $R_{21}$ do not change (less than 10\%) at
different resolutions, thus we adopt the average value as the estimate
of the ratio at $14''$, which is listed in Table 5.

The $R_{10}$, $R_{21}$ ratios were taken from Braine et al. (2003) with a
resolution of $11''$.  We have re-examined the data and estimate
that the lower limit for $R_{21}$  is 30 at the bridge assuming a 2
$\sigma$ upper limit for the $^{13}$CO(1--0) intensity.

It is well known that interacting galaxies have an unusually high
$^{12}$CO/$^{13}$CO line intensity ratio $R_{10}$ (Aalto et al. 1991,
1995, 1997; Casoli et al. 1991; Casoli, Dupraz \& Combes 1992a, 1992b;
Henkel \& Mauersberger 1993; Henkel et al. 1998).  In the Taffy
galaxies, the high ratios of $R_{10}$ and $R_{21}$ were seen only in
the bridge, which is similar to the case in the Antennae galaxies in
which a high ratio of $R_{21}$ was seen only in the overlap
region. These extreme values can provide strong constraints on the
physical parameters of the molecular clouds using an excitation
analysis model.

\subsection{Excitation Analysis}

  To estimate
 the physical parameters of the molecular gas, we have
 employed a large velocity gradient (LVG) model (\eg Goldreich \&
 Kwan 1974), in which it is assumed that the systematic
 motions rather than the local thermal velocities dominate the
 observed linewidths of the molecular clouds.
We use
 this model to fit the observed line ratios in Table 5 for different
 combinations of (\Tk, n(H$_2$), $\Lambda $), where \Tk  is the kinetic
 temperature and $\rm \Lambda =Z_{CO}/(dV/dR)$, with $Z_{CO}$=[$
 ^{12}$CO/H$_2$] being the fractional abundance of $^{12}$CO with respect to
 H$_2$ and dV/dR the velocity gradient. The optimum set of parameters
 is determined by  minimizing $\chi ^2$.

As outlined in Zhu et al. (2003), a grid
of LVG models was searched in the parameter range of
$\rm T_{\rm kin} =5-200$ K, $\rm
n(H_2)=(0.1-10^4)\times 10^3$ \cm3,
$\Lambda = (0.1-10^2)\times 10^{-6}$ \XdVunit,
and $\eta$ = \1213CO = 20 -- 200.
These ranges cover all possible conditions found in the GMCs of our
Galaxy as well as in external galaxies
(see the discussion in Zhu et al. 2003 and Yao et al. 2003).

Three positions including the bridge, the central region of UGC 12915 and
the southern disk of UGC 12914,  have been chosen for model fitting.
The results are summarized in Table 6.  As pointed out by Yao et al. (2003),
the CO-to-H$_2$ conversion factor $X$ can be directly derived from the LVG
parameters using the formula
\begin{equation}
X= \frac{n(H_2) \Lambda }{ Z_{co} T_{rad}}
\end{equation}
Where $T_{rad}$ is the radiation temperature for the $^{12}$CO(1--0) line
transition (Yao et al. 2003).

We found that the line ratios in all these regions can be fitted
reasonably well using a single component model. It is well
understood that models incorporating more data for starburst
galaxies require at least two components -- a prevailing optically
thin, warm, diffuse component and an optically thick, cool, dense
component with a smaller filling factor (e.g., Aalto et al. 1995).
However, comparisons between single and  double-component models
yield approximately the same column density of H$_2$, which is
weighted strongly toward the prevailing diffuse component (e.g., Zhu
et al. 2003). Therefore, we will use a single component model which
requires less free parameters. The results for different regions are
analyzed in the following sections.

\subsubsection{The Disks of UGC 12915 and UGC 12914}

The BIMA map of CO(1--0) shows that the molecular gas in the central
region of UGC 12915 is distributed in a circumnuclear ring, this is
concurred by higher resolution OVRO CO(1--0) data (Iono et al. 2005).
The central $14''$  corresponds to the inner 4 Kpc
which covers the majority of this ring.  In the southern disk of UGC
12914 (UGC 12914S), the region for LVG modeling is on the disk $ \sim
17'' - 18''$ or 5 kpc away from the nucleus. Thus the gas properties
should be comparable to that of a galaxy disk rather than a starburst
nucleus.

The best fit parameters for UGC 12915 correspond to
\Tk= 15 K,
${{\rm n(H_2)}}$ = $ 3.1 \times 10 ^3$ \cm3,
$\Lambda $ = $  4.0\times 10^{-6}$ \XdVunit
which yields
$r_{21}$=0.80,
$r_{31}$=0.45,
$R_{10}$= 12,
$R_{21}$=17
(see Fig. 9a).
Other combinations of parameters can also fit the ratios within the
observational uncertainties.  In general, $\eta =40-70$, \Tk $=15-35 K
$, \nh2$ = 800-10000$ \cm3, $\Lambda = 1- 25 \times 10^{-6} $ \XdVunit
can fit the ratios with $\chi ^2 < 2$. Outside these ranges it is
still possible to fit the observed line ratios, but those models would
require an impossibly low CO abundance or unacceptably high velocity
gradients.  The range in the second line in Table 6 is our best
estimate of the range of parameters with reasonable physical
conditions.

The ranges of \nh2 and $\Lambda$ span one order of magnitude.  But the
column density $N_{CO}/\Delta V$ = \nh2 $ \Lambda $ can be constrained
in a narrow range $2.9 - 6.1 \times 10^{19}$ cm$^{-2}$ (km/s)$^{-1}$.
According to formula (2), the consequence of a well confined
$N_{CO}/\Delta V$ is that the $X$ factor
can also be constrained in a narrow range $2.2-6.1 \times 10^{19}$
($\frac{Z_{CO}}{10^{-4}} $) \Xunit.  $Z_{CO}$ is generally assumed to
be in the range $10^{-5} -10^{-4}$ (see the discussion in Zhu et
al. 2003).  We use the average value $5 \times 10^{-5}$ here, bearing
in mind the large uncertainty in this parameter and therefore in the
derived $X$ factor.  Our results indicate that $X$ is 2 -- 6 times
(assuming ${Z_{CO}}=0.5 \times 10^{-4}$) lower than the conventional
value of $X=2.8 \times 10^{20}$ \Xunit (Bloemen et al. 1986; Strong et
al.  1988), but comparable to $X=0.5 \times 10^{20}$ \Xunit estimated
from the diffuse clouds in the Galaxy by Polk et al. (1988), and with
that found in nearby starbursts galaxies (e.g., M82, Mao et al. 2000),
interacting galaxies (Zhu et al. 2003) and IRAS luminous galaxies (Yao et
al. 2003).  Downes and Solomon (1998) used a model of radiative
transfer through subthermally excited CO in a molecular disk to
derive the gas mass in the central region of ultra-luminous galaxies,
and they also found a factor of 5 less molecular gas mass than that
derived from the standard X factor.

The southern disk of UGC 12914 has similar observed line ratios,
except that the $^{13}$CO(2--1) data are unavailable. The best fit
LVG parameters (Table 6) are similar to those of UGC 12915, but the
average gas density appears to be lower. The $X$ factor is in the
range $4.6-9.0 \times 10^{19}$ \Xunit, closer to the conventional
value.  The molecular clouds in both galaxies appear to remain cold,
which could be due to a low level of star formation.

Braine et al. (2003) suggested that the molecular gas mass is
over-estimated by a factor of 4 in UGC 12915, but not over-estimated
in UGC 12914 (they used $X=2.0 \times 10^{20}$ \Xunit).  This is
consistent with our results considering the large uncertainty in
$Z_{CO}$.  The northern disk of UGC 12914 has weaker CO emission, a
lower dust temperature and less star formation activity compared to
the southern disk (UGC 12914S), thus the $X$ factor in the northern
disk could be higher than that of UGC 12914S and closer to the
conventional value.

\subsubsection{The Bridge}

In the bridge, the high $R_{10}$ value puts a strong constraint on the
[$^{12}$CO]/[$^{13}$CO] abundance ratio $\eta$. In LTE cases, assuming
$^{13}$CO is optically thin, we have $R_{10} = (1-e^{-\tau})
\eta/\tau$, or $\eta = R_{10} \tau /(1-e^{-\tau})$. Thus the high
value for $R_{10}$ implies either very high $\eta$ or very low optical
depth. If the molecular gas clouds are similar to GMCs in
our Galaxy ($\tau > 0.5$), $\eta$ must be greater than 56 to
satisfy $R_{10}$ = 43. For a given $R_{10}$, a lower value of $\eta$ means a
lower optical depth of CO(1--0).

For non-LTE cases, our LVG modeling requires $\eta > 50$ in order to
fit the high $R_{10}$ ratio of 43. For example, when $\eta=50$, the
best fit parameters require $\tau_{10}$=0.2 -- 0.3, which is lower
than that of most GMCs in our Galaxy or other galaxies. Only when
$\eta \ge 60$ do the model fits yield an acceptable value of $\tau >
0.5$. Table 6 lists the best estimates of parameters with $\eta = 70$
(Br1 in Table 6), which yields the ratios
$r_{21}$=0.88,
$r_{31}$=0.46,
$R_{10}$= 38,
$R_{21}$=44.
The $\chi^2$ contours are shown in Fig. 9(b). 

Higher $\eta$ can always fit the observed ratios equally well, for
example the set of parameters Br2 in Table 6 have $\eta =100$, \Tk=30 K,
\nh2$=1500 $ \cm3 , $\Lambda= 4.7 \times 10^{-6} $ \XdVunit, can
produce the ratios $r_{21}$, $r_{31}$, $R_{10}$, $R_{21}$ = 0.84, 0.46,
41, 50. Such a model does not require optically thin conditions (in this
case $\tau_{10}$ = 1.3).

As reviewed by Wilson \& Rood (1994), $\eta$ is normally found to
be in the range 40--75 in our Galaxy, with lower $^{13}$CO abundances 
found in the nuclear region of starburst galaxies (e.g., Henkel \&
Mauersberger 1993). There is only one report of  $\eta > 100 $  in
the literature, in which  $\eta= 150 \pm 27 $ was found in the
diffuse clouds of $\zeta$ Ophiuchi (Sheffer et al. 1992). In the
Taffy bridge, an LVG solution with a high $\eta$ is attractive
because it does not require optically thin $^{12}$CO, and the velocity
gradient is more reasonable. The solution with $\eta=70$ requires
$\Lambda=   2.7 \times 10^{-6}  $ \XdVunit, which implies dV/dR = 18
km/s pc$^{-1}$ if $Z_{CO} = 5 \times 10^{-5}$, comparing to dV/dR =
10 km/s pc$^{-1}$ when $\eta=100$. The velocity gradients in the
molecular clouds in our Galaxy are around 1 km/s pc$^{-1}$ (Bonnel
et al. 2006; Knapp et al. 1988; Larson 1981). It is
possible that dV/dR is higher in the bridge of the Taffy galaxies
due to shocks and strong turbulence, but a significantly higher
velocity gradient would be difficult to explain. Therefore, we could
not rule out an extremely high $\eta$ in the bridge. In \S 4.2,
we discuss  some scenarios that can reduce the $^{13}$CO abundance.

As listed in Table 6, we can constrain the LVG parameters in the range
\nh2$=(1.0 - 3.1) \times 10^3$ , \Tk = 20--50 K,
$\Lambda = (1 - 4.7) \times 10^{-6}$ \XdVunit,
$\eta$ =  60--100.
$N_{CO}/\Delta V$ = $(0.7 - 2.2) \times 10^{16} $ cm$^{-2}$ (km/s)$^{-1}$.
The range of $X$ can be constrained in a narrow range,
$X= (2 - 3.6) \times 10^{19}$ cm$^{-2}$ (K km/s)$^{-1}$, which is
lower than that of the  disks of UGC 12915 and UGC 12914,
mainly due to the unusually high $R_{10}$ and $R_{21}$.
Besides the abundance $\eta$, the molecular clouds in the
bridge appear to have a higher \Tk, lower optical depth and
higher velocity dispersion.

Comparing the Taffy bridge LVG parameters and $X$ factor with
the overlap region of the Antennae galaxies N4038/9 (Wilson et
al. 2000, 2003; Zhu et al. 2003), the ISM in the Taffy bridge seems to
be more diffuse, with a lower optical depth and smaller $X$
factor. Similarly, Jarrett et al. (1999) have shown that the warm
dust in the bridge is similar to that of high latitude cirrus clouds in our
Galaxy. Hence the molecular clouds in the Taffy bridge could be
similar to the Galactic diffuse clouds.  The bulk of ISM in the overlap region of the
Antennae is between the Taffy disk ISM and the bridge diffuse clouds.

\subsection{Molecular Gas Mass and SFE }

Table 7 lists the mass of the molecular gas
derived using the new $X$ factors. From Table 6 we can see
that the $X$ factor is similar for both disks, thus we
used the average
$X=7.8 \times 10^{19}$  $\rm cm^{-2}$ \rm (K km $\rm s^{-1})^{-1}$,
which is 4 times lower than the conventional
$X$ factor $X = 2.8  \times 10^{20}$ $\rm cm^{-2}$ \rm (K km $\rm
s^{-1})^{-1}$. As a result, the molecular gas mass is 4 times lower
than that estimated by GZS03. In the bridge, $X=2.6
\times 10^{19} $  $\rm cm^{-2}$ \rm (K km $\rm s^{-1})^{-1}$
 and thus $M_{H_2}$
is reduced to 10 times less than the estimate in GZS03.
For comparison we also list the dust mass, \ion{H}{1} mass, \lir and the
derived ratios.  The definition of the bridge region is somewhat
arbitrary but to be consistent with previous studies we used a region
similar to that of GZS03.

Our new result indicates that the global \McoHI ratio is approximately
0.6, which is 4 times higher than the mean  \McoHI ratio for an
optically selected sample of late-type galaxies (0.15) by Boselli et al.
(2002). UGC 12915 has a \McoHI ratio similar to UGC 12914, while in the
bridge the \McoHI ratio is a factor of 4 lower than
that in the UGC 12915 disk.

The global SFE is 5.2 $\ls/\ms$ in UGC 12915 which is similar to that
of normal spiral galaxies and one order of magnitude lower than that
from IR luminous galaxies in the SLUGS survey (Yao et al. 2003).
This is consistent with the low $r_{31}$ ratio found in this system. The
SFE is twice as high in UGC 12915 compared to UGC 12914.

We cannot fit an SED to the bridge region as there are no resolved FIR
fluxes. To estimate the dust mass in the bridge we have noted that the
flux ratio $S_{450}/S_{850}$ is very similar for the bridge and UGC
12915 and so we have assumed that the cold dust temperature is also
similar. Thus we have scaled the dust mass for the bridge from that of
UGC 12915 by the ratio of their 850$\mu$m fluxes, producing a bridge
dust mass of $1\times 10^7 \ms$. A high resolution FIR map of the
system (e.g. with Spitzer or Herschel) would be required to produce a more
accurate estimate of the dust masses and also the SFE for the bridge
region. The global
gas-to-dust ratio in the two galactic disks is consistent with that of
our Galaxy, but is a factor of 2--3 higher in the bridge. 

The high flux ratio $S_{450}/S_{850}$ in the bridge (as high as that
in UGC 12195) suggests that (for $\beta=2$) the dust is not colder in
the bridge region despite being further from the stellar disks (which
based on geometric factors should lead to a temperature about 67\% of
that in the disks). This is interesting in light of the very distinct
properties of the bridge molecular gas and may be related. However,
given the lack of FIR data from which to constrain the temperatures,
this is not yet strong evidence for a source of heating within the
bridge.

\subsection{A Gas-to-dust Ratio Map}

In order to examine the variation of the gas-to-dust ratio across the
system, we have convolved the CO and SCUBA 850 $\mu$m maps to match
the resolution of that of \ion{H}{1} ($18''$) and combined them to
derive the gas-to-dust ratio map.

The \ion{H}{1},  H$_2$ gas mass and dust mass can be estimated by using the
following formula:
\begin{equation}
M_{\rm H I} (M_{\sun}) = 2.36 \times 10^5 D^2 S_{\rm HI}
\end{equation}
\begin{equation}
M_{\rm H_2}(M_{\sun}) = 2.2 \times 10^3 D^2 S_{\rm CO}
\end{equation}
\begin{equation}
M_{\rm dust}(M_{\sun}) =\frac{S_{850} D^2}{\kappa _{850} B_\nu(T_{\rm d})}
\end{equation}
where $D$ is luminosity distance in Mpc, $ S_{\rm HI}$ and $S_{\rm
CO}$ is the \ion{H}{1} and H$_2$ integrated flux in Jy km $\rm
s^{-1}$, $S_{850}$ is the flux at 850 $\mu$m (after correcting for the
CO(3--2) contamination), $B_\nu(T_{\rm d})$ is the Planck function, and
$\kappa _{850}$ is the dust opacity per unit dust mass which is
assumed to be = $0.77 \,{\rm cm^2 g^{-1}}$ (c.f., James et al. 2002; Dunne
et al.  2000). The dust temperature $T_{\rm d}$ was taken to be 21 K
which is the value from fitting the SED to the total fluxes in \S 3.2.

To derive \Mco using formula (4) we used $X $ = $5.6 \times 10^{19}$
$\rm cm^{-2}$ \rm (K km $\rm s^{-1})^{-1}$ which is the average value
from our LVG modeling. We apply a single $X$ factor to the whole map
because the division of different CO regions was not well defined and
we do not want to introduce artificial variation in the map. Readers
should bear in mind that the molecular gas mass could be slightly
underestimated in the two disks, but overestimated in the bridge because
a different scaling factor was used. However, this difference has
little effect on the total gas-to-dust ratio as \ion{H}{1} 
dominates the gas mass in the bridge.

After applying the proper scaling factors, the 850 $\mu$m continuum,
and the \ion{H}{1} and H$_2$ integrated intensity maps are essentially
the mass distribution of different ISM components.  Using the AIPS
task COMB, we first added the CO and \ion{H}{1} map together to make a
map of the total gas mass $[M(H_2)+M(HI)] \propto (2.2 \times 10^3
S_{\rm CO} + 2.36 \times 10^5 S_{\rm HI})$, and then divided it by the
850 $\mu$m continuum map. The result is a map of the gas-to-dust ratio
= $[M(H_2)+M(HI)]/M_{dust}$, which is shown in Fig 10. This map
clearly shows that the gas-to-dust ratio has little change within the
two disks, and in both galaxies this ratio is close to that of our
Galaxy (200). However, the gas-to-dust ratio is a factor of
approximately 3 times higher in the bridge. As discussed in \S 3.5,
the $X$ factor in the bridge is very likely to be lower than the value
used for scaling the map ($ 5.6 \times 10^{19}$ $\rm cm^{-2}$ \rm (K
km $\rm s^{-1})^{-1}$), hence the gas-to-dust ratio could be lower
than that shown in the map.

For comparison, we have also divided the CO(1--0) map by the 850
$\mu$m continuum to make a map of the $M_{H_2}/M_{\rm dust}$
ratio. Interestingly, we found little variation in the $M_{H_2}/M_{\rm
dust}$ ratio across the whole system. This $M_{H_2}/M_{\rm dust}$
ratio map is not shown here because it is featureless.  This suggests
that the dust distribution follows the molecular gas closely, and the
high total gas-to-dust ratio in the bridge is mainly due to the high
\ion{H}{1}/dust ratio.

\section{Discussion}

\subsection {Origin of the molecular gas and dust in the bridge }

The bridge region of the Taffy galaxies is remarkable in that it
contains a large amount of atomic and molecular gas as well as cold
dust. Our LVG modeling shows that the molecular gas is optically
thin in the bridge region, and the $X$ factor is 10 times lower than
the conventional value. The \Mco is reduced to $1.3 \times 10^9
\ms$, which is only 22\% of the total atomic gas mass in the same
region. Thus the atomic gas is the major contributor to the gas mass
of the bridge. 

There are two plausible origins for the molecular gas seen in the
Taffy bridge, which we will investigate in more detail below:
\begin{enumerate}
\item{The gas originates in GMCs in the parent galaxies. The GMCs collide during the interaction and their momentum is canceled leaving them in the region between the two disks. The molecules are dissociated by shocks during the collision but are able to reform afterward.} 
\item{The large amount of \ion{H}{1} gas in the bridge used to be part of the diffuse phase of the ISM in the disks of the galaxies. It was removed from the disks during the collision and is now being converted into \H2.}
\end{enumerate}

\subsubsection{A GMC origin for the molecular gas}

Our LVG analysis shows that the brightness dilution factor=
$\frac{T_{\rm mb}}{T_{\rm rad}} = (\frac{\Delta v} {\Delta V}) f_{\rm
vol}$ is $\sim 0.02$ for the molecular clouds in both galaxy
disks. The velocity filling factor $(\frac{\Delta v} {\Delta V})$ is
$\sim 0.1$ considering the typical velocity dispersion $\Delta v = 20$
\kms in individual clouds and the FWHM line width $\Delta V = 200 $
\kms, thus the volume filling factor $f_{\rm vol}$ is 0.2.  Using the
formula in section \S 4.3 in Braine et al. (2004), we estimate that
the surface filling factor is $\sim 0.8$, which is much higher than
the threshold value 0.2 and thus GMC-GMC collision is inevitable.

During the collision, extremely dense regions (e.g. protostars) will
act as collisionless bodies like stars and will not affected by the
ram pressure. Otherwise we would expect to see more stars in the
bridge, which is not the case from the Spitzer IRAC images at 3.5-5.8
$\mu$m.  The collision is most likely to occur between the diffuse low
density gas surrounding the high density cores.  A higher density
clump will be more likely to remain in the galaxy disk as it will
possess a greater momentum (which would need to be canceled in order
for it to fall out of the disk). The \ion{H}{2} region outside the
disk of UGC 12915 appears to be a dense region that lost a small
amount of momentum which caused it to slow down gradually and
eventually drop out of the disk.  Such morphology was confirmed in
both the CO and sub-mm continuum map. This is not due to a gas
excitation effect, but rather reflects a density gradient in the
bridge that is consistent with the cloud collision scenario.

The large amount of extra-disk molecular gas (even
after reduced by a factor of 8 due to the $X$ factor) was initially a
puzzle for Condon et al. (1993) because any molecules involved in
collisions that are strong enough to blow a molecular cloud out of a
disk would be shock destroyed in the process, and the conversion of
\ion{H}{1} to \H2 is too slow under normal circumstances to have
replenished the $H_2$ outside the disks (Braine et al. 2003).

However, as pointed out by Harwit et al. (1987) and Braine et al.
(2003, 2004), while GMC-GMC collisions result in complete ionization
at a temperature of $T > 10^6$ K, the cooling time for dense gas from
$10^6$ K to $10^4$ K is less than 100 years, and \H2 can form quickly
in the dense post shock region. The recent theoretical work by Bergin
et al. (2004) provides a more detailed picture of molecular cloud
formation behind shock waves. They show that it takes $\sim 10^6$ yr
for cooling from $10^4$ K to 25 K, and the H$_2$ formation time scale
is $\sim 4 \times 10^6$ yr at the density of $n = 2500$
\cm3 and T= 23 K , which is much shorter than the age (20 Myr) of
the Taffy bridge (Condon et al. 1993).  Hence, as pointed out by Gao
et al. (GZS03), direct GMC-GMC collision is a possible mechanism to
produce large amount of molecular gas in the Taffy bridge.

This scenario is supported by the similar $M_{\rm H_2}/M_{\rm dust}$
ratio across the disks and the bridge suggesting a common
origin. However, the shocks which are responsible for destroying the
molecules post-collision would also be expected to have an effect on
the dust grains. The fact that the bridge has a normal 
$M_{\rm H_2}/M_{\rm dust}$ ratio indicates that either the dust is not
efficiently destroyed by the shocks or is able to reform {\em in
situ\/} on the time-scale of 20 Myr. Dust is thought to be able to
grow in the ISM by coagulation (following grain-grain collisions) or
by accreting gas particles. Both processes occur more quickly in dense
environments and also require the existence of seed nuclei to begin
with. Thus complete destruction of grains into the gas-phase by shocks
is unlikely in this case as there is then no way for the grains to
re-grow on these time-scales. If the grains were sputtered or broken
up into smaller fragments by the shock then is it possible that they
could regrow on these time-scales? According to Whittet (2003) the
time-scale for grain coagulation is given by

\[t_c = \frac{3.2\times 10^{8} \rm{yr}}{n_H v_d}\]
\noindent
where $t_c$ is the collision time-scale in yrs, $n_H$ is the gas
density in $cm^{-3}$ and $v_d$ is the relative velocity between grains
in $km s^{-1}$. Using a typical value of $v_d \sim 0.1 \, km s^{-1}$
for clouds (Jura 1980) and the value of $n_H$ from our LVG model
($3000 \,cm^{-3}$), we find $t_c \sim 1$ Myr. This is much less than
the time since the collision (20 Myr) so it is possible that dust can
have re-grown in this way.

The time-scale for accretion of atoms from the gas-phase into mantles
is given by Jones (2005) as $10^9 \rm{yr} / n_H \sim  30$ Myr for the
bridge clouds. Thus growth by accretion is not likely to be a fast
enough mechanism for replenishing the dust population in the bridge.

Given that coagulation does not change the overall mass in dust, but
rather changes the size distribution this does not help us recover a
normal gas-to-dust ratio if there was total destruction of grains during
the collisions. An investigation of the dust properties (composition
and size distribution) in the bridge and disks using Spitzer IRS would
provide useful constraints on the possible history of the bridge dust and thus the origin of the molecular gas.  

Finally, the apparent offset between the CO, dust and \ion{H}{1} peaks in
the bridge (Figs 4 and 5) also reflects the collision scenario: if the
\ion{H}{1} peak marks roughly the point of impact, the CO and dust peaks
in the bridge imply the dragging of the disk clouds of UGC 12915 from
the point of impact with dense molecular gas being the most difficult to
detach from the disk. 

\subsubsection{Formation of molecular gas from the bridge \ion{H}{1}}

An alternative hypothesis is that the molecular gas is forming from
the large reservoir of \ion{H}{1} in the bridge. The conversion of
\ion{H}{1} to \H2 is too slow under normal circumstances to have
accounted for the $H_2$ outside the disks (Braine et
al. 2003). However, recent simulations by Glover \& Mac Low (2006)
suggest that H$_2$ forms over 1-2 Myr in turbulent gas, much faster
than the gravitational free-fall timescales typically required to form
H$_2$ in gravitationally unstable, initially static gas. The large
velocity gradients fitted to the bridge region would suggest a large
degree of turbulence could be present and so this scenario also seems
to be plausible. However, we would expect a certain degree of spatial
correlation between the \ion{H}{1} and CO distributions in this case,
which is not seen in our data. Furthermore, it is difficult to explain
the abnormally high gas-to-dust ratios seen in the bridge whereas a
normal Galactic $M_{\rm H_2}/M_{\rm dust}$, comparable to that of both
disks, is observed for the bridge molecular component. The general
dust content of the atomic gas from which the \H2 is purportedly
forming is low compared to the disks of the galaxies, $M_{HI}/M_{\rm
dust}$ in the bridge is 600 compared to 139 and 174 in UGC 12194/5
respectively. The dust deficient atomic gas must then somehow
produce a molecular component which has a gas-to-dust ratio comparable
to the clouds in the disks. This would require some fine-tuning of the
atomic-molecular conversion process and so we feel it is less
attractive. However, we cannot rule it out without further
detailed modeling which is beyond the scope of this paper.

\subsection {Gas properties in the bridge }

Our LVG modeling indicates that the molecular properties of
the bridge are similar to that of the warm, diffuse ISM in our Galaxy,
with a low optical depth, low average gas density and
a gas temperature higher than that of the galactic disks.
The $X$ factor is similar to that in the FIR cirrus clouds
(Knapp \& Bowers 1988).
The $^{12}$CO/$^{13}$CO abundance ratio $\eta$ is probably higher than
that in the Galactic centre, which is also a characteristic for
diffuse clouds.

There are three mechanisms that can possibly explain the high
[$^{12}$CO]/[$^{13}$CO] ratio:

1) Sheffer et al. (1992) suggested that selective isotopic
photo-dissociation can reduce the $^{13}$CO abundance in diffuse
cloud. Since CO is photo-dissociated by UV radiation, $^{13}$CO is
destroyed preferentially because $^{12}$CO is self-shielded to a
greater extent.  They reported that $\eta =150 \pm 27 $ was observed
in the warm diffuse clouds in $\zeta $ Ophiuchi. However, there is no
observational evidence for massive star formation in the bridge and so
no source of UV photons to dissociate the $^{13}$CO.

2) Henkel \& Mauersberger (1993) suggested that $^{13}$CO is
deficient in the outer parts of galactic disks because
$^{13}$C is produced by old stars (e.g., red giants)
whereas more $^{12}$CO is produced by young massive stars. Regions with more
massive star formation would have a higher $^{12}$CO abundance
while the outer part of a galaxy and high latitude clouds would have less
$^{13}$CO. If the gas from the bridge is mainly from the northern disk
of UGC 12914, it would contain less $^{13}$CO.

3) The destruction and recombination of molecules after shocks could
change the [$^{12}$CO]/[$^{13}$CO] abundance ratio.  Although it is possible
    that H$_2$ and $^{12}$CO molecules can form quickly after shocks, the
    formation of $^{13}$CO is slower than $^{12}$CO because it is not
    self-shielded and thus needs a higher gas and dust density to
    prevent photo-dissociation by UV radiation. Also the rare isotope
    molecules like $^{13}$CO would take longer time to form as
    collisions between $^{13}$C and dust grains are less frequent. Thus
    the time scale $2 \times 10^7$ years after
    the collision might not be long enough for $^{13}$CO to
    reach an equilibrium abundance,  and as a result the
    [$^{12}$CO]/[$^{13}$CO] ratio is higher than normal.

Unfortunately, detailed chemical evolution models that take account of
the formation of all molecules are not available. Observationally, we
can not distinguish between the above three scenarios, although we
favour 2 and 3 due to the lack of an obvious source of UV photons.

If the [$^{12}$CO]/[$^{13}$CO] ratio in the bridge is indeed higher,
the molecular gas would have a higher optical depth, smaller velocity
gradient and a slightly higher $X$ factor. A sample solution with
$\eta =100$ is listed in Table 6.  Nevertheless, even in the extreme
case with $\eta =200$, the $X$ factor $= 3.3 \times 10^{19}$ \Xunit in
the bridge is still less than that in the two disks and the overall
gas-to-dust ratio is still higher than the Galactic value.

There are some other possible mechanisms that could affect the gas
properties in the bridge. The direct \ion{H}{1} -- \ion{H}{1} cloud
collision speed could be as high as 900 km/s, since the relative speed
in the counter-rotating gas disks is over 600 km/s and the transverse
velocity of the collision of the two disks is nearly 700 km/s (Condon
et al. 1993). Such a high speed collision will inevitably create high
temperatures and associated X-ray emission. Cold dust would not easily
survive in this hostile environment unless the gas condenses enough to
partially shield itself from the intense hot
radiation.\footnote{However, sputtering lifetimes will depend strongly
on the density of the X-ray plasma and so sensitive X-ray observations
with XMM or Chandra would be required to check this scenario} This
could explain why the $M_{HI}/M_{\rm dust}$ ratio is higher than the
$M_{H_2}/M_{\rm dust}$ ratio in the bridge, e.g., most dust related to
the more diffuse
\ion{H}{1} clouds has been destroyed.

In addition, supersonic speed can induce large scale shocks as
evidenced here in the expanding ring structures in the bridge where
stars are forming (though at low level in the Taffy system compared to more
extreme cases in classical ring galaxies such as Cartwheel). Weak
ring patterns are also evident in PAHs (Fig. 2) which are part of
the features mixed with the more bridge-like morphologies in CO and
dust in the bridge. In short, besides the main bridge molecular
features extending further from the huge CO concentration in the \ion{H}{2}
regions, part of molecular gas in the bridge could have originated
from other mechanisms (e.g., rings) and have different gas properties
as well.

\subsection {Large velocity gradients}

Our LVG model assumes a velocity gradient higher than that found in
normal molecular clouds in our Galaxy.  Assuming $Z_{CO} = 5 \times
10^{-5}$, which is the value observed in GMCs in our Galaxy (Blake et
al. 1987), the velocity gradients are dV/dR = 12 km/s pc$^{-1}$ in UGC
12915 and 10 km/s pc$^{-1}$ at the bridge according to the values of
$\Lambda $ in Table 6.

As discussed in Zhu et al. (2001, 2003) and Yao et al. (2003), the
velocity gradient under the virial condition would be around 1--2
km/s pc$^{-1}$ for a spherical cloud with a mean gas density of $1.5
\times 10^3$ \cm3. Thus most the clouds in our LVG models are 
gravitationally unbound. In star forming regions in our Galaxy,
outflows can easily drive the gas velocity to dV/dR $\sim 20$ km/s
pc$^{-1}$ (e.g. Moriarty-Schieven et al. 2006), and it is possible
that part of the velocity gradient in UGC 12915 is associated with
star formation sites.  For the Taffy bridge, however, no obvious
star formation is seen outside the \ion{H}{2} region in the highly
sensitive Spitzer image at 8 $\mu$m (Fig. 2), which rules out the
hypothesis of large scale inflows and outflows. Hence the velocity
gradient should be from strong turbulence in the diffuse clouds.

As pointed out by Condon et al.(1993) the energy stored in the magnetic
field of the bridge exceeds the turbulent energy in the gas. If there is
equipartition between magnetic field and cosmic ray particle energy, then
the sum of the two, which is a minimum value, may act to resupply the
turbulent energy, offsetting the effects of dissipation by radiation from
shocks. It is thus plausible that the inferred velocity gradients are
associated, not with molecular clouds, but with transient density enhancements
driven by turbulence at least partially resupplied by magnetic field and
cosmic rays.

On a related point, the cosmic ray particles associated with the
excess synchrotron emission originating in the bridge may have their
origin partly in the shocks associated with the collision, though
some fraction must also originate from a pre-existing supply in the
disks. Assuming equipartition, the minimum particle and magnetic
energies are each about $4 \times 10^{55}$ erg (Condon et al. 1993),
a very small fraction (0.2\%) of the total gas kinetic energy
associated with the collision which is at least $2 \times 10^{58}$
erg. Thus the role of the fields and particles might be to ``store
collision energy'', acting perhaps as a long lived supply for the
observed turbulence.


\subsection {Low SFE in the gas rich interacting pair?}

Our revised global SFE is now comparable to that of normal spirals,
and lower than most of starburst galaxies and IR-luminous galaxies though
higher than that estimated by Gao et al. (GZS03) due to the factor of
3--4 lower molecular gas mass used here. The SFE in UGC 12915 is 40\%
higher than in UGC 12914. Judging from the SFE, UGC 12915 now
appears to be a starburst.  However, it should be noted that most of
the SFE values published in the literature used the conventional $X$
factor to estimate the molecular gas which might underestimate the
SFE. Comparing with the average SFE for IRAS bright galaxies in Yao
et al. (2003), in which the $X$ factor was also derived from a LVG
analysis, UGC 12915 has a much lower SFE and thus does not qualify
for a starburst.

A larger discrepancy now exists in the bridge because there is not a
clear indicator that best describes the local star formation rate in
this region. GZS03 interpret the 20cm-to-CO ratio as a SFE indicator
because the radio continuum emission has been widely used as a tracer
of the recent star formation in galaxies (\eg, Condon \etal 1996;
Condon 1992). Furthermore, the tight correlation between radio
continuum and CO (GZS03) serves as a good argument to suggest that the
local star formation rate can be approximated by radio continuum
though reduced by some large factor.  Using this method GZS03 found
the SFE in the bridge to be mostly comparable to that of UGC
12914. Now, with the large reduction in the molecular gas mass in the
bridge and assuming that the star formation rate can still be
indicated by the radio continuum emission, the resulting SFE in the
bridge will be a factor of 3 higher than that of UGC 12914, which is
not consistent with the MIR data.  Thus, most of radio continuum
emission from the bridge cannot have originated from local star
formation and so cannot be used as a local star formation
tracer. Condon et al. (1993) estimated that synchrotron emission from
radiative electrons in the Taffy bridge has enhanced the radio
continuum emission by more than 50\%. If the star formation rate
estimated from the radio continuum is reduced by a factor of 2-3, the
new SFE in the bridge would be again comparable to that of the UGC
12914.

Since the 8 $\mu$m is virtually optically thin, the Spitzer 8 $\mu$m
data clearly show little massive star formation in the bridge except
for the \ion{H}{2} region and along the faint ring structures of UGC
12914. However, were PAHs destroyed by shocks and intense hot X-ray
radiation? Appleton et al. (2006) found abundant H$_2$ molecules in
the large scale bow shocks in the intergroup medium outside the
galaxies in the Stephan Quintet, and star formation is obviously
on-going as evidenced by the strong H$_2$ emission and \ha~ emission,
yet no PAHs were detected. PAH emission is usually a good indicator of
star formation in galactic disks, but it gets fainter in outer disks
and may not be able to trace star formation in regions like the Taffy
bridge. The star formation rate estimated from IRAC 8$\mu$m PAH
emission may therefore be underestimated except for the \ion{H}{2} regions. 

It should be noted that the combination of the \ion{H}{1} and H$_2$ (even
with a conversion factor $X$ nearly ten times smaller than the
conventional one as deduced here) still yields a gas surface density
higher than the threshold for star formation in normal galactic
disks (Martin \& Kennicutt 2001). Therefore, stars could be forming
in the bridge, although it is also questionable that whether the
criteria for disk star formation could be applied to the special
geometry like the Taffy bridge. The sensitivity constraints of the
KAO far-IR observations nevertheless set an upper limit to the star
formation rate in the bridge to be less than $\sim$20\% of that from
the disks of the galaxies. We will have to wait for Spitzer's sensitive
MIPS observations to better estimate the local star formation rate in
the bridge and possibly resolve these issues.


\subsection {Comparison with numerical simulations}

The Taffy system is particularly unique as a laboratory for testing
some numerical simulations. With a nearly face-on, counter-rotating
collision, the relative collision speed of HI clouds is close to
900\kms~ as the estimated transverse collision speed of the two
galaxies is also about 600--700\kms (Condon et al. 1993). At such a high
speed, the diffuse gas clouds (mainly HI) collide at highly supersonic
speeds.  Thus, we expect an extensive ``gas splash'', and possibly
other observable effects of shock heating and radiative cooling as
probed by numerical simulations (Struck 1997; Barnes 2002, 2004).

A simple estimate of the amount of HI left in the taffy bridge was attempted by
Gao et al. (2003). For an ideal face-on disk collision with the same
counter-rotating speed, all HI clouds in two disks that have collided
inelastically will be left behind after collision, canceling all
systematic velocities of the HI clouds, particularly if the transverse
collision velocity is much smaller than the rotation speed. When the
collision speed is much higher than the rotation speed, the amount of
gas that collides and gets left behind will be less, depending upon
the exact area of the disk-to-disk overlap and the amount of area of
the disks that has been swept owing to the counter-rotation during
their close impact. The Taffy system falls between the two extreme
cases as the collision speed is comparable to the counter-rotation
speed. We indeed observe here that nearly half of the total HI is
located between the two stellar disks whereas only less than a quarter
of molecular gas is left in the taffy bridge. This is because HI
clouds have a disk area filling factor which is essentially unity, an
order of magnitude larger than the disk area filling factor of the
molecular gas. The amount of molecular gas left behind after collision
is mostly controlled by this disk area filling factor besides other
factors such as the rotation and collision speeds and the disk overlap
area etc.

According to Toomre \& Toomre (1972), the impact parameter should be
smaller than the disk radii in order to produce the ring structures,
rather than the prominent tidal tails. The Taffy system has both ring
structures and weak tidal features, therefore the impact parameter
cannot be much smaller than the disk radii since numerical simulations
show that encounters with an impact parameter larger than the galactic
radii can be most effective in producing long tails and stronger gas
inflow toward the nuclei.  Given these considerations, most of the southeast
portion of the target disk (UGC 12914) might be much less damaged. A
small portion of the intruder's (UGC12915) HI disk in the northwest may have
suffered less impact as well, owing to this particular impact
configuration. Judging from the HI distribution in Fig.~5, we can
assume that more than 1/3 of HI clouds in the target galaxy reside in
the southeast disk and about 1/3 of HI gas in the intruder galaxy is located
in its northwest disk. Then most of the HI in the intruder and the HI in the 
northwest
portion of the target galaxy, totaling about half of the total HI gas
of the system, could have experienced a collision, owing to the
sweeping of the counter-rotating disks, with the remnants mostly left
behind in the taffy bridge.

Most numerical simulations of galaxy interactions (e.g.  Barnes \&
Hernquist 1996; Springel et al. 2005; Di Matteo et al. 2005) focused
on merging systems and tidal interactions, and could not be applied to
the Taffy system which is a transient event.  Struck (1997) simulated
the direct collisions between two gas rich disks with heating and
cooling. His models show that face-on disk-disk impacts are highly
dissipative, and a large amount of gas can be splashed out into the
bridge between the galaxies. The splash bridge is a transient event
and the gas will eventually be accreted back onto the parent
galaxies. Shock heating is strong on the impact, but there is rapid
expansion cooling in the bridge with a time scale of $10^7$ yr.
Virtually no star formation in the ``splash'' bridge was predicted by
these models because the gas density is low there.  All these results
are qualitatively consistent with observational data, in particular
the HI. However, the models of Struck (1997) could not explain the
existence of a large amount of molecular gas in the Taffy bridge
because large clouds are expected to be dispersed by the impact.  In a
more recent work, Struck and Smith (2003) modeled the starburst
collisional pair NGC 7714/15 which have an optical morphology similar
to that of the Taffy system. These simulations can reproduce the star
formation history in the bridge as well as in the disks.  Apparently,
the model results depend on the assumption of the star formation law
and are sensitive to the details of the interactions (Barnes 2004;
Struck 2005). The Taffy system is colliding at a speed much higher
than that of the NGC7714/15 pair (900 km/s vs 380 km/s) with different
mass ratios and disk orientations, thus a different detailed
simulation is needed to predict the fate of the bridge gas and its
capability of forming stars. Exploring different numerical simulations
to match the observational details is beyond the scope of this paper
and will be part of our future work.

It is also worth noting that a substantial population of luminous IR
galaxies (LIRGs) with dust much cooler than the local LIRGs may exist
at moderate/high redshifts (Chapman et al. 2002).  There is also some
indication that there may be a high redshift population of sub-mm
sources with excess radio emission compared to the IR/radio
correlation for local galaxies.  Taffy is indeed at a rather cold dust
temperature and does have excess radio continuum emission in the
bridge and perhaps the conditions required to produce a taffy-type
system may have occurred more often at high redshift where collisions
between gas-rich systems are much more frequent.


\section{CONCLUSIONS}

We have obtained highly sensitive sub-mm maps of the Taffy system at
450 and 850 $\mu$m.  The galaxies are detected at 450 $\mu$m for the
first time, and both wavelengths show the presence
of cold dust in the bridge connecting the two galaxies.  CO(3--2) line
emission was detected in both disks of UGC 12914 and UGC 12915 as well
as in the bridge. Our conclusions are
summarized as the following:

1. The CO(3--2)/(1--0) integrated intensity ratio is  in the range 0.3
-- 0.4 throughout the system and it shows little variation. This is
consistent with rather cold dust temperature and low average gas density
in the Taffy system.

2. The gas-to-dust ratio is comparable to the Galactic value in the two
disks, but is a factor of $\sim 3$ higher in the bridge. However, the
CO/dust mass ratio is not higher in the bridge, and the high gas-to-dust
ratio is due to the high $M_{HI}/M_{\rm dust}$ ratio. The
$M_{HI}/M_{H_2}$ ratio is close to unity in the two disks and a factor
of 3--6 times higher in the bridge, thus the dust appears to be most
closely related to the molecular component in the bridge.

3.  Our LVG excitation analysis suggests that the $X$ factor in the
two galaxy disks is $ \sim 7.8 \times 10^{19} $ \Xunit, 3--4 times
lower than the standard Galactic value, and the total amount of
molecular gas in the Taffy galaxies is 3--4 times lower than previous
estimates based upon the standard $X$ factor.

4. The bridge has an extremely high $R_{21}$ and $R_{10}$ 
($R = I(^{12}CO)/I(^{13}CO)$) ratios, suggesting
that the $^{13}$CO abundance might be lower than that of the two
galaxy disks. The molecular gas has a low optical depth and its
physical condition is comparable to that in the diffuse clouds in
our Galaxy. The new estimate of the molecular gas mass is $ \sim 1.3
\times 10^9 \ms$. The majority of the gas in the bridge is in atomic
gas phase.

5. Our result is consistent with the scenario that the large amount of
 molecular gas in the bridge is originated from the disk molecular
 clouds and is a result of momentum canceling in the GMCs due to
 direct GMC-GMC collisions. The physical condition of the molecular
 gas in the bridge is comparable to that in the diffuse clouds in our
 Galaxy. The gas might have been ionized by strong shocks, but has
 quickly cooled down and re-formed molecules. The normal CO/dust ratio
 in the bridge shows that grain destruction in the shocks is either
 not effective, or that the grains have managed to reform.

6. Little starburst is found in the Taffy system except for the intruder
galaxy UGC 12915. The Spitzer 8$\mu$m map shows little star 
formation activity in the
bridge except in the \ion{H}{2} region. The global SFE is comparable to
that of normal spiral galaxies.

\acknowledgements

The authors wish to thank the staff of the JCMT for their generous
assistance.  We also thank Tom Jarrett, Jim Condon for providing
digital images. YG acknowledges the supports from NSF of
China (distinguished young scholars and grant \#10333060) \& Chinese
Academy of Sciences (hundred-talent).  ERS acknowledges the support of
a Discovery Grant from the Natural Sciences and Engineering Research
Council of Canada.

\clearpage
{\bf Figures:}

\figcaption{ The 850$\mu$m contours overlaid on an 450$\mu$m signal-noise map.
Contour levels start at 2$\sigma$ and increment at $\sqrt{2}$. Both
maps have been smoothed slightly, and the circles at the lower left
indicate the smoothed telescope beams ($9.4''$ at 450$\mu$m and $16''$
at 850$\mu$m). The rms is 3.3 mJy beam$^{-1}$ and 11.8 mJy beam$^{-1}$
at 850/450$\mu$m
\label{fig-1}}

\figcaption{ The 450 $\mu$m contours overlaid on a near-IR
 image from the Spitzer IRAC at 8 $\mu$m.
The contour levels start at 2$\sigma$ and increment by $\sqrt{2}$.
The circle at the lower left indicates the smoothed telescope beam ($9.4''$).
The rms is $1\sigma = 11.8$ mJy.
\label{fig-2}}

\figcaption{ The 450 $\mu$m contours overlaid on a mid-IR  image at 15 $\mu$m
from the  ISO  (Jarrett et al. 1999). The circle at the lower left indicates the smoothed telescope beam ($9.4''$).
\label{fig-3}}

\figcaption{(a) The 450$\mu$m and (b) 850 $\mu$m contours
overlaid on a CO(1--0) image from the BIMA interferometer.
Contour levels start at 2$\sigma$ and increment at $\sqrt{2}$, and
the circles at the lower left indicate the smoothed telescope 
beams ($9.4''$ at 450$\mu$m and $16''$ at 850$\mu$m).
\label{fig-4}}

\figcaption{ The 850 $\mu$m contours
overlaid on an  \ion{H}{1} image from the VLA.
Contour levels are the same as in Fig 4b.
\label{fig-5}}

\figcaption{  SED fitting of the total dust emission in  
UGC 12914/15 using a (a) single-component model; (b) two-component
model; (c) two-component model for UGC 12915; (d) two-component model
for UGC 12914. The filled circles with error bars are the data points
at 25, 60, 100, 450 and 850 $\mu$m. Note that the 25$\mu$m point is not
'fitted' but used as an upper constraint to the temperature of the
warm component.
\label{fig-6}}

\figcaption{ The CO 3--2) profiles overlaid
on the BIMA CO(1--0) zero-momentum color map.
\label{fig-7}}

\figcaption{ The JCMT CO(3--2) profiles overlaid on the BIMA CO(1--0) 
profiles (from Gao et al. 2003) at the three positions listed in Table 5.
The CO(1--0) data have been convolved to 14$''$ to match 
the resolution of the CO(3--2) data.
\label{fig-8}}

\figcaption{ The $\chi^2$ contours (in logarithm) from the LVG
modeling for (a) UGC~12915 (left) at the best fit temperature $T_{\rm
K} =15$ K with $\eta=60$;  
(b) the bridge (right) at the best fit temperature $T_{\rm K} =35$ K 
with $\eta=70$.
\label{fig-9}}

\figcaption{  A color map of the  gas-to-dust ratio
$[M(H_2)+M(HI)]/M_{\rm dust}$.
The contours are the 850 $\mu$m dust emission with levels
at 0.01, 0.014 ... 0.056 Jy/beam (increment $\sqrt{2}$).
\label{fig-10}}

\clearpage

\newpage
Fig 1 see fig1.gif \\

Fig 2 see fig2.gif \\

Fig 3 see fig3.gif \\

Fig 4 see fig4a.gif and fig4b.gif \\

Fig 5 see fig5.gif \\

\newpage
Fig 6. 
\includegraphics[angle=-90,scale=0.3]{fig6a.ps}
\includegraphics[angle=-90,scale=0.3]{fig6b.ps}
\includegraphics[angle=-90,scale=0.3]{fig6c.ps}
\includegraphics[angle=-90,scale=0.3]{fig6d.ps}
\newpage
Fig 7 see fig7.gif \\
\\
\newpage
Fig 8
\includegraphics[scale=0.7]{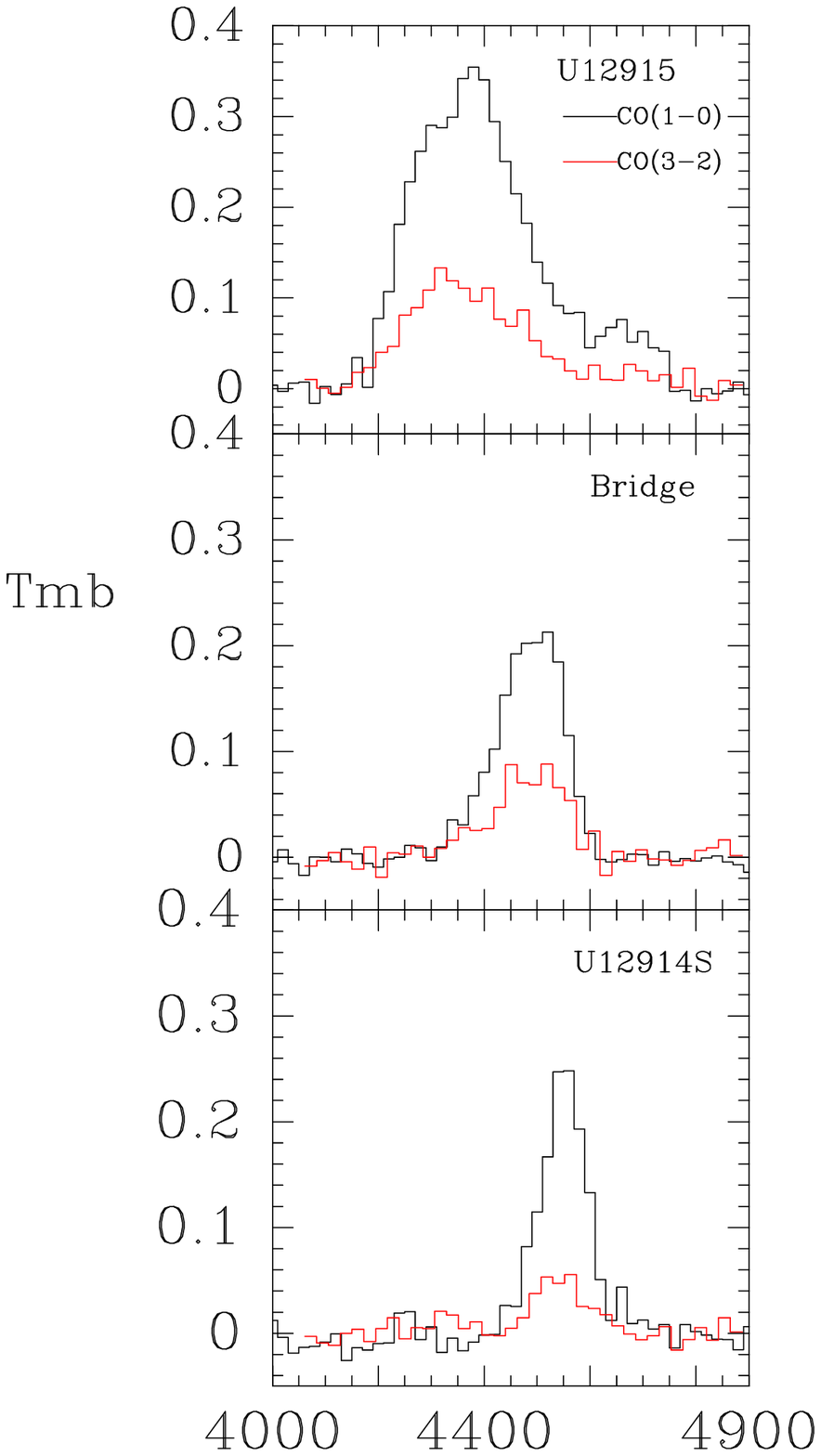}
\newpage
Fig 9
{\plottwo{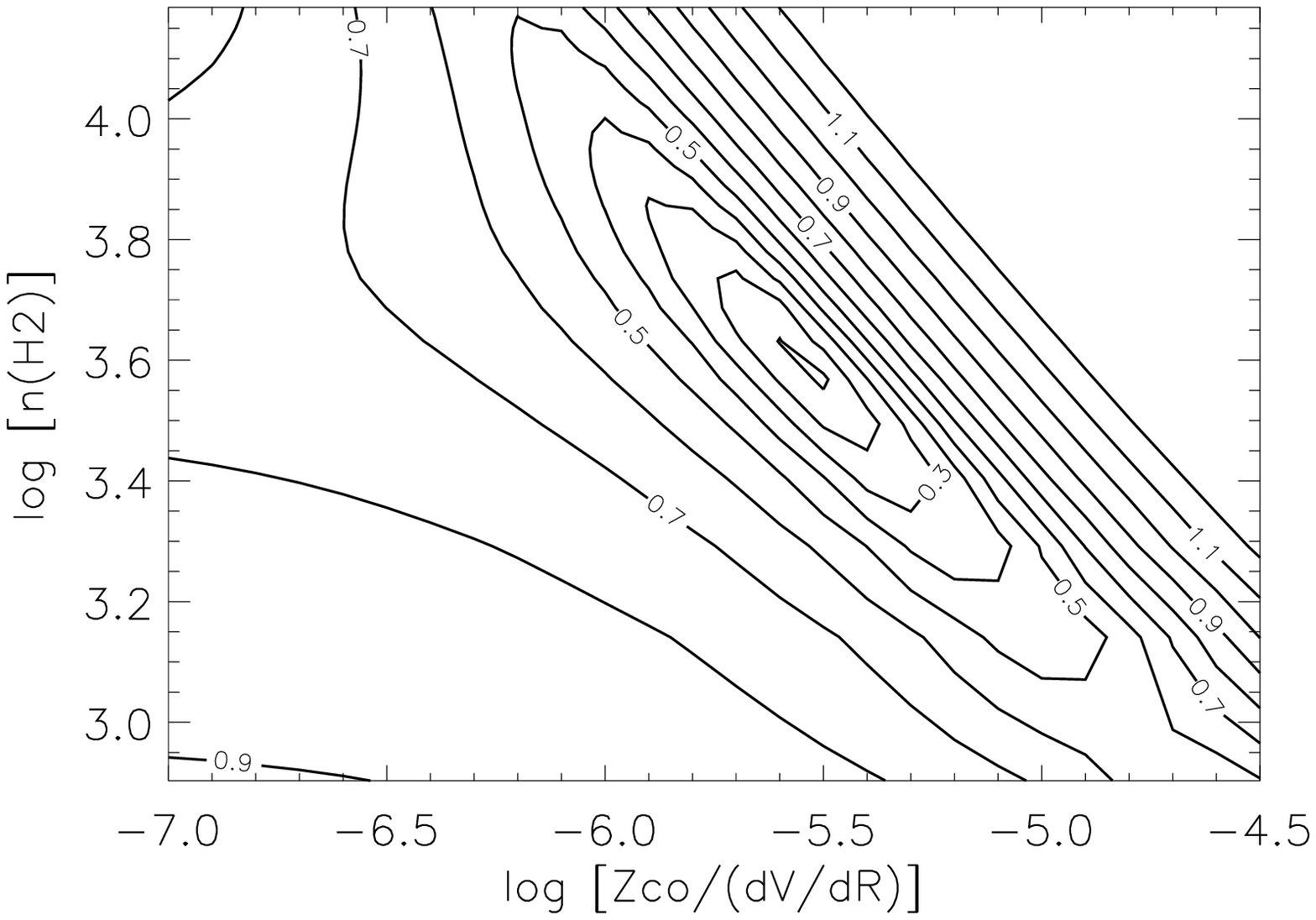}{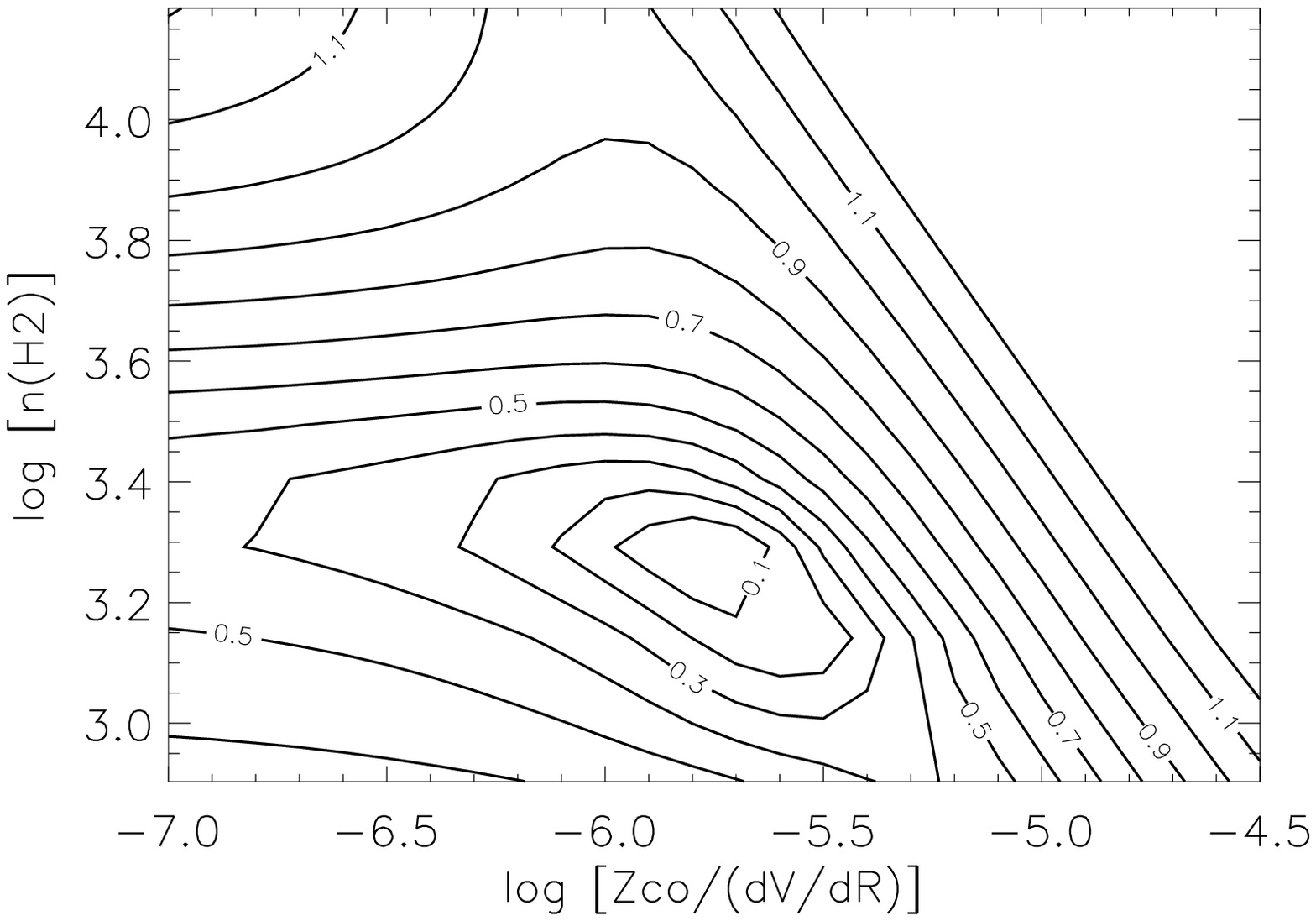}}
\newpage
Fig 10
\plotone{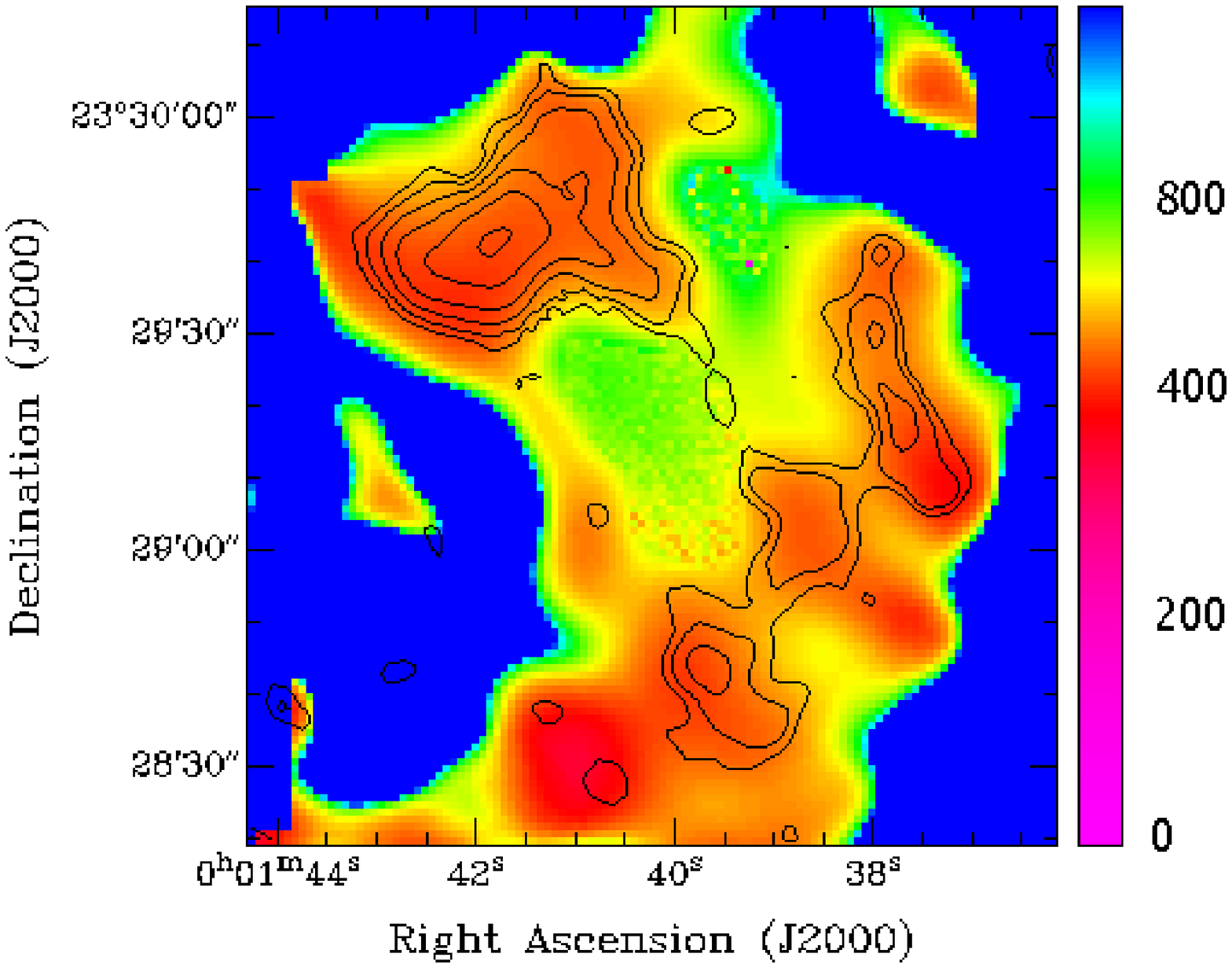}

\clearpage

{\small
\begin{center} 
{ \bf Table 1. Obs details}
              
\vskip 2.ex
\tabcolsep 6pt
\begin{tabular}{lccccccc}
\hline
\hline
{Obs. Date} &
{Calibrator} &
{FCF(450$\mu$m) $^a$} &
{FCF(850$\mu$m) $^a$} &
{Chop} &
{CSO $\tau$} &
{Int. Time} 
\\
\hline
20040916  &   Uranus   &   401   &     285 & $150''$ & 0.030 &  2.5 Hr\\
20021223  &   CRL618   &   263   &     213 & $150''$ & 0.027 &  2.5 Hr\\
19980710  &   Uranus   &   945   &     255 & $120''$ & 0.058 &  1.2 Hr\\
19980421  &   Uranus   &   901   &     240 & $120''$ & 0.071 &  1.3 Hr\\
\hline
\end{tabular}
\end{center}
\noindent
$ ^{\rm  a}$  Column 3 and 4 are the Flux conversion factor derived using
the flux calibrator in column 2.
\\
\vspace*{0.3cm}
}
\clearpage

{\small
\begin{center} 
{\bf \large Table 2: Integrated intensity ratios of $^{12}$CO J=3--2/1--0 }
\vskip 2.ex
\tabcolsep 16pt
\begin{tabular}{ccccccc}
\hline
\hline
   $\Delta \alpha     \, \, ^a $  &
   $\Delta \delta      \, \, ^a $    &
$I_{32} \, \, ^b $  &
$\sigma_{32} \, \, ^c$  &
$ I_{10} \, \, ^d$ &
$r_{31} \, \, ^e$  &
$\sigma_{31} \, \, ^f$ 
\\
( $''$     ) &
( $''$   ) &
(K \kmps) &
(K \kmps) &
(Jy \kmps) &
          &
\\
\hline
0&0    &      20.5  &   2    &    151   &            0.42  &  0.05 \\
-4.5&2 &      22    &   1    &    163   &            0.41  &  0.05 \\
-7&0   &      16    &   2.2  &    135   &            0.37  &  0.05 \\
0&-7   &      11.2  &   1.2  &    65    &            0.48  &  0.05 \\
-7&14  &      9.2   &   1.5  &    66    &            0.43  &  0.06 \\
-14&14 &      9.0   &   0.5  &    74    &            0.38  &  0.03\\
-7&7   &      18.6  &   2    &    130   &            0.45  &  0.04\\
7&0    &      7.7   &   2.2  &    80    &            0.30  &  0.05 \\
7&7    &      5.4   &   0.5  &    49    &            0.34  &  0.06\\
0&7    &      14.9  &   1    &    110   &            0.42  &  0.06\\
-22&-22 &     4.2    &  1.1  &    21    &            0.55  &  0.2 \\
-11&-11 &     6.7    &  1    &    44    &            0.47  &  0.06\\
7&-7  &       8.5    &  2.   &    48    &            0.54  &  0.2 \\
-7&-7 &       $<$8.0   &  -    &    51    &            $>$0.52 &  -   \\
-14&7 &       9.9    &  1.3  &     95   &            0.33  &  0.1\\
-14&0 &       9.4    &  1.   &    78    &            0.38  &  0.1 \\
-22&-9 &    8.4   &   0.8 &     59   &             0.44 &   0.05 \\
-22& -16  &    7.3   &   0.8 &     43   &             0.52 &   0.05 \\
U12914S&  &    4.5   &   0.5 &     46   &             0.31 &   0.05 \\
\hline
\end{tabular}
\end{center}
\begin{small}
\noindent
\noindent
$ ^{\rm a}$   $\Delta \alpha$ and   $\Delta \delta$  
are the R.A and DEC offsets from the map center R.A.= $23^h59^m08.4^s$, DEC= $23^o13'01''$ (B1950). 
$ ^{\rm b}$ The integrated intensity ($ T_{\rm mb}$) of CO (3--2).
$ ^{\rm c}$ The  uncertainty in $I_{32} $.
$ ^{\rm d}$ The integrated intensity ($ T_{\rm mb}$) of CO (1--0) in  a 14$''$ beam.
$ ^{\rm e}$ The ratio of  $I_{32}/I_{10}$.
$ ^{\rm f}$ The uncertainty in the $r_{31}$ ratio.
\end{small}
}

\clearpage

{\small
\begin{center} 
{ \bf Table 3: Comparison of submm fluxes with other wavelengths}
              
\vskip 2.ex
\tabcolsep 6pt
\begin{tabular}{lccccccccc}
\hline
\hline
{Position $^a$ } &
{$S_{850} $ $^b$ } &
{$S_{450} $ $^c$} &
{$\frac{S_{450}}{S_{850}} $ } &
{$S_{15} $$^d$ } &
{$S_{\rm CO}$ $^e$} &
{$S_{\rm HI}$  $^f$} &
{$\frac{S_{CO}}{S_{850}}$}  &
{$\frac{S_{CO}}{S_{450}}$}  &
{$\frac{S_{CO}}{S_{HI}}$  } 
\\
{ } &
{(\rm mJy) } &
{(\rm mJy) } &
{ } &
{(\rm mJy) } &
{( \rm Jy K km s$^{-1}$) } &
{( \rm Jy K km s$^{-1}$) } &
{     } &
{     } &
{     } 
\\
\hline
U12915	&109&	921	&8.4    & 276	& 355 & 5.0   &  3.3 & 0.39 & 71\\
U12914	&108 &	757	&7.0    & 168	& 315 & 4.9   &  2.9  & 0.42 & 64\\
Bridge  &48 &   406     &8.5    & 8.2 	& 100 & 1.4   &  2.1  & 0.25 & 71\\
U12915C	&77 &	674	&8.8    & 	& 213 & 1.1   &  2.8  & 0.32 & 194\\
U12914S	&32 &	289	&9.0    & 	& 92  & 1.0   &  2.9  & 0.32 & 92\\
U12914C	&27 &	244	&9.0	&       & 88 &  0.90  &  3.3  & 0.36 & 98\\
U12914N	&24 &	201	&8.4    &    	& 81  & 0.89  &  3.4  & 0.40 & 91\\
\hline
\end{tabular}
\end{center}
\noindent
$ ^{\rm a }$ The first three rows are the global values for each
galaxy and the bridge region.  The remaining rows are for a $25''$
region centering at the nuclei of the two galaxies (C) and the north
and south MIR knots in the disk of UGC12914.  $ ^{\rm b }$ The
measured 850 $\mu$m flux density without correcting for CO(3-2)
contamination.  $ ^{\rm c }$ The 450 $\mu$m flux density. $ ^{\rm d }$
The ISO 15 $\mu$m flux density from Jarrett et al. (1999).  $ ^{\rm e
}$ The CO(1--0) flux density from the BIMA map (GZS03).  $ ^{\rm f }$
The HI flux density from Condon et al. (1993).


}


{\small
\begin{center} 
{ \bf Table 4: Dust Mass from SED fitting using a two component model}
              
\vskip 2.ex
\tabcolsep 6pt
\begin{tabular}{lcccccccc}
\hline
\hline
{Position} &
{$S_{850}$ $^a$} &
{$S_{450}$ $^a$} &
{$S_{100} $$^b$ } &
{$S_{60} $ $^c$} &
{$T_{\rm c}$ $^d$} &
{$M_{\rm c}$  $^d$} &
{$T_{\rm w}$ $^d$} &
{$M_{\rm w}$  $^d$} 
\\
{ } &
{(\rm Jy)} &
{(\rm Jy)} &
{(\rm Jy)} &
{(\rm Jy)} &
{($ K $)} &
{($ 10^7 \ms$)} &
{($ K $)} &
{($ 10^7 \ms$)} 
\\
\hline
U12914$^{\rm e}$  & 0.085  &0.96  & 3.51   & 2.1   & 18  & 4.3 &  43 &  0.03\\
U12915$ ^{\rm  e}$   & 0.087  &1.124 & 10.6  & 4.31   & 24  & 2.8 &  42 &  0.056\\
Total $^{\rm  f}$    & 0.172   &2.084 & 14.1  & 6.4   & 21  & 6.8 &  42 &  0.09\\
\hline
\end{tabular}
\end{center}
\noindent
$ ^{\rm a}$ The SCUBA flux density at 850 and 450 $\mu$m. The CO(3-2)
contamination has been removed from the 850 $\mu$m fluxes.  $ ^{\rm
b}$ The FIR flux density at 100 $\mu$m. The total flux is from IRAS
(Sanders et al. 2003), the flux for UGC12915 is from KAO. The flux for
UGC12914 is derived by subtracting the UGC12915 flux from the total
flux.
$ ^{\rm c}$ The FIR flux density at 60 $\mu$m. The ratio of flux between the two galaxies was taken from the HIRES paper by Surace et
al. (2004) but the values scaled to the total flux from Sanders et al. (2003)  
$ ^{\rm d}$ $T_{\rm c}$, $T_{\rm w}$, $M_{\rm c}$ and $M_{\rm
w}$ are the temperature and dust mass of the cold and warm component
from the SED fitting using a two-component model (see text for
details).  
$ ^{\rm e}$ Note, the bridge is not resolved by IRAS and
its flux will be inextricably contained in the quoted fluxes for
UGC12914 and UGC12915. We will therefore allocate the submm bridge flux equally between the two galaxies also.

$ ^{\rm f}$ The total fluxes of whole system, including the
bridge and two disks. The total dust mass is derived by fitting the
SED of the total fluxes.

}
\clearpage

{\small
\begin{center}
{\large \bf Table 5: Observed line ratios $^a$}
\vskip 2.ex
\tabcolsep 16pt
\begin{tabular}{lccccc}
\hline
\hline
Position  &
$r_{21}$  &
$r_{31}$  &
$R_{10}$  &
$R_{21}$  \\
\hline
U12915   &  0.75 $\pm$ 0.1  &0.42 $\pm$ 0.1  &   14 $\pm$  3  & 15 $\pm$ 3 \\
Bridge   &  0.79 $\pm$ 0.15  &0.44 $\pm$ 0.1  &   43   &  $> 30$  \\
U12914S$ ^{\rm b}$
         &  0.69 $\pm$ 0.15  & 0.30 $\pm$ 0.1  &  14.3 $\pm$  4 &  -- \\
\hline
\end{tabular}
\end{center}

\noindent

\noindent
$ ^{\rm  a}$ The average line intensity ratios in  a 14$''$ beam. 
The intensities with a higher resolution have been convolved
to 14$''$.
$ ^{\rm b}$ UGC12914S is at offset ($-44''$,$-55''$) from UGC12915 nucleus.
 The CO(2--1) ratios are from the average of JCMT and 
IRAM data with $11''$ and $20''$ resolutions.

}



{\small
\begin{center} 
{\large \bf Table 6: One-component model fitting result}
\vskip 2.ex
\tabcolsep 3pt
\begin{tabular}{lcccccccc}
\hline
\hline
{Position} &
{\nh2} &
{\Tk} &
{$Z_{CO}$/(dV/dR)} &
$N_{CO}/\Delta V$ &
$\tau_{10}$  &
$X$             &
{[$\frac{^{12}CO}{^{13}CO}$]} &
{$\chi^2   $} 
\\
{} &
{$10^3$ \cm3} &
{K} &
{$ 10^{-6}$ pc (km/s)$^{-1}$} &
{$10^{16}$ cm$^{-2}$ (km/s)$^{-1}$} &
{ } &
{ $10^{19}$ } &
{} &
{}
\\ 
\hline  
U12915   &  3.1       &  15     &    4.0  & 3.89     &  3.5      &  7.8     &   60     &   1.3\\ 
         &  0.8-- 10  &  10--40 &  1-- 25 & 3--6     &  1.8--5.1 & 4.4--12.2 &   40--70 &   $<$2 \\
 
Br1      &  1.5       &  35     &    2.7  &  1.25    &  0.7      & 2.6     &  70      &   1.2 \\
Br2      &  1.5       &  30     &    4.7  &  2.17    &  1.3      & 3.6      &  100      &  0.7 \\
         & 1-- 3.1    & 20--50  &  1--4.7 &0.7--2.2 & 0.4--1.3  & 2.0--3.6 &  60-100  &   $<$2 \\
 
U12914S  &  1         &  15    &   12.7    & 3.9      &   5.1     & 9.0      &   70     &   0.6\\ 
         &  0.5-- 2.5 &  10--45 & 4.7-- 19& 2.6--4.5 & 2.3--4.1  &4.6--10.0  &   40--70 &   $<$2 \\ 
\hline 
\end{tabular}
\end{center}

\noindent
Columns 2--4 contain the LVG parameters noted above, column 5
contains the CO column density, column 6 the optical depth at the J=1--0
transition;
column 7 contains the $X$ factor in the unit of 
cm$^{-2}$ (K km\,$s^{-1}$)$^{-1}$, 
 assuming $Z_{CO}= 0.5 \times 10^{-4}$.
column 8 contains the assumed isotope abundance ratio, and 
column 9 shows the value of $ \chi^2 $ associated with the fit.   
Note: for each position, the first row is the best fit parameters and
the second row is the estimated range within the observed uncertainty.
\vspace*{2cm}
}

\clearpage

{\small
\begin{center} 
{ \bf Table 7: Masses of different components, FIR luminosity and their ratios}
              
\vskip 2.ex
\tabcolsep 6pt
\begin{tabular}{lccccccc}
\hline
\hline
{Position} &
{$F_{CO}$ $^a$} &
{$M(H_2)$ $^b$} &
{$M_{dust}$  $^c$} &
{$M_{HI} $ $^d$} &
{$L_{IR} $ $^e$}&
{$\frac{M_{H_2}+M_{HI}}{M_{Dust}}$ } &
{$L_{IR}/M_{H_2} $ }  
\\
{ } &
{\rm Jy Km s$^{-1}$} &
{$ 10^9 \ms$} &
{$ 10^7 \ms$} &
{$ 10^9 \ms$} &
{$ 10^{10} \lsun$} &
    &
{$\lsun/\ms$ }
\\
\hline
U12915   &  355   & 4.2    &    2.3   &       4   &   3.6    & 356   & 8.6  \\
U12914   &  315   & 4.2    &    3.6   &       5   &   1.4    & 255   & 3.4 \\
Bridge   &  325   & 1.3   &     1.0   &       6   &          & 730   &     \\
Total    &  995   & 9.7   &     6.9   &       15  &   5      & 358   & 5.2  \\
\hline
\end{tabular}
\end{center}
\noindent
$ ^{\rm  a}$  The CO(1--0) flux from Table 1 of GZS03. 
\\
$ ^{\rm  b}$  The molecular gas mass derived using from the $X$ factor
in Table 6. 
\\
$ ^{\rm c}$ The dust mass estimated by fitting the SED for each region
separately. In this case the submillimetre fluxes for UGC 12914 and
12915 do not include the bridge region. However, the bridge is not
resolved in the FIR and so an SED fit is not possible. The dust mass
for the bridge was estimated assuming the cold temperature is similar
to that in UGC 12915 (see text for details).
\\
$ ^{\rm d}$ The atomic gas mass derived from the HI VLA map (GZS03).
\\
$ ^{\rm e}$ The FIR luminosity from IRAS and KAO (Zink et al. 2000), see 
text for details
\\
\vspace*{0.3cm}
}

\end{document}